\documentclass[letterpaper,twocolumn,10pt]{article}
\usepackage{usenix,epsfig,endnotes}

\usepackage{graphicx}
\usepackage{usenix,epsfig,endnotes}
\usepackage{amsmath}
\usepackage[numbers]{natbib}

\usepackage{booktabs} 
\usepackage[utf8]{inputenc}
\usepackage{verbatim}
\usepackage{subcaption}
\usepackage{cleveref}
\usepackage{enumitem}
\usepackage[printwatermark]{xwatermark}
\usepackage{color}
\usepackage{fancyvrb}

\newcommand{\System}[0]{{Ray}}
\newcommand{\system}{ray}

\newcommand{\secref}[1]{Section~\ref{#1}}

\newcommand{\figref}[1]{Figure~\ref{#1}}
\renewcommand{\eqref}[1]{Equation~\ref{#1}}

\newcommand{\tabref}[1]{Table~\ref{#1}}

\usepackage{microtype}
\everypar{\looseness=-1}
\begin{document}

\date{}

\title{\Large \bf Ray: A Distributed Framework for Emerging AI Applications}

\author{
{\rm Philipp Moritz\footnotemark[1],\, Robert Nishihara\footnotemark[1],\, Stephanie Wang, Alexey Tumanov, Richard Liaw,}\\
{\rm Eric Liang, Melih Elibol, Zongheng Yang, William Paul, Michael I.~Jordan, Ion Stoica}\\
University of California, Berkeley
} 

\maketitle

\thispagestyle{empty}

\renewcommand{\thefootnote}{\fnsymbol{footnote}}
\footnotetext[1]{equal contribution}

\subsection*{Abstract}
The next generation of AI applications will continuously interact with
the environment and learn from these interactions.
These applications impose new and demanding systems requirements, both in terms
of performance and flexibility. In this paper, we consider these requirements
and present \System{}---a distributed system to address them. \System{}
implements a unified interface that can express both task-parallel and
actor-based computations, supported by a single dynamic execution engine. To
meet the performance requirements, {\System} employs a distributed scheduler and
a distributed and fault-tolerant store to manage the system's control state. In our
experiments, we demonstrate scaling beyond 1.8~million tasks per second and
better performance than existing specialized systems for several challenging
reinforcement learning applications.

\input{fig/pygment.tex}

\section{Introduction}
\label{sec:intro}

Over the past two decades, many organizations have been collecting---and
aiming to exploit---ever-growing quantities of data.  This has led to
the development of a plethora of frameworks for distributed data
analysis, including batch~\citep{mapreduce,spark-nsdi12,dryad},
streaming~\citep{flink,naiad,storm}, and graph~\citep{graphlab,pregel,graphx}
processing systems.  The success of these frameworks has made it possible
for organizations to analyze large data sets as a core part
of their business or scientific strategy, and has ushered in the age of ``Big Data.\!''

More recently, the scope of data-focused applications has expanded to
encompass more complex artificial intelligence (AI) or machine learning (ML)
techniques~\citep{jordan2015machine}.  The paradigm case is that of
\emph{supervised learning}, where data points are accompanied by
labels, and where the workhorse technology for mapping data points
to labels
is provided by deep neural networks.  The complexity of these deep
networks has led to another flurry of frameworks that
focus on the training of deep neural
networks and their use in prediction.  These frameworks often
leverage specialized hardware (e.g., GPUs and TPUs), with the goal
of reducing training time in a batch setting.  Examples include
TensorFlow~\citep{tensorflow-osdi16}, MXNet~\citep{mxnet-learningsys},
and PyTorch~\citep{pytorch}.

The promise of AI is, however, far broader than classical supervised learning.
Emerging AI applications must increasingly operate in dynamic environments,
react to changes in the environment, and take sequences of actions to
accomplish long-term goals~\citep{agarwal2016multiworld,nishihara2017real}.  They must
aim not only to exploit the data gathered, but also to
explore the space of possible actions.  These broader requirements are
naturally framed within the paradigm of \emph{reinforcement learning} (RL).
RL deals with learning to operate continuously within an uncertain environment
based on delayed and limited feedback~\citep{sutton1998reinforcement}.  RL-based
systems have already yielded remarkable results, such as Google's AlphaGo
beating a human world champion \citep{silver2016mastering}, and
are beginning to find their way into dialogue systems, UAVs
\citep{ng2006autonomous}, and robotic manipulation~\citep{GuHolLilLev17,
van2010superhuman}.

The central goal of an RL application is to learn a policy---a mapping from
the state of the environment to a choice of action---that yields effective
performance over time, e.g., winning a game or piloting a drone.  Finding effective policies in large-scale applications
requires three main capabilities.  First, RL methods often rely on \emph{simulation}
to evaluate policies.  Simulations make it possible to explore many different choices
of action sequences and to learn about the long-term consequences of those choices.
Second, like their supervised learning counterparts, RL algorithms
need to perform \emph{distributed training} to improve the policy based on
data generated through simulations or interactions with the physical
environment.  Third, policies are intended to provide solutions to control
problems, and thus it is necessary to \emph{serve} the policy in interactive
closed-loop and open-loop control scenarios.

These characteristics drive new systems requirements: a system for RL
must support \emph{fine-grained} computations (e.g., rendering actions
in milliseconds when interacting with the real world, and performing
vast numbers of simulations), must support \emph{heterogeneity} both in
time (e.g., a simulation may take milliseconds or hours) and in resource
usage (e.g., GPUs for training and CPUs for simulations), and must support
\emph{dynamic} execution, as results of simulations or interactions
with the environment can change future computations. Thus, we
need a dynamic computation framework that handles millions
of heterogeneous tasks per second at millisecond-level latencies.

\looseness=-1
Existing frameworks that have been developed for Big Data workloads
or for supervised learning workloads fall short of satisfying these
new requirements for RL. Bulk-synchronous parallel systems such
as MapReduce~\cite{mapreduce}, Apache Spark~\cite{spark-nsdi12}, and
Dryad~\cite{dryad} do not support fine-grained simulation or policy serving.
Task-parallel systems such as  CIEL~\cite{ciel}  and
Dask~\cite{dask-scipy15} provide little support for distributed training and
serving. The same is true for streaming systems such as Naiad~\cite{naiad} and
Storm~\citep{storm}. Distributed deep-learning frameworks such as
TensorFlow~\cite{tensorflow-osdi16} and MXNet~\citep{mxnet-learningsys} do not
naturally support simulation and serving. Finally, model-serving systems such
as TensorFlow Serving~\citep{tensorflow-serving} and Clipper~\cite{clipper}
support neither training nor simulation.

While in principle one could develop
an end-to-end solution by stitching together several existing systems (e.g.,
Horovod~\cite{horovod} for distributed training, Clipper~\cite{clipper} for
serving, and CIEL~\cite{ciel} for simulation), in practice this approach is
untenable due to the {\em tight coupling} of these components within applications.
As a result, researchers and practitioners today build one-off
systems for specialized RL applications~\cite{tian2017elf, gorila, silver2016mastering, dota, salimans2017evolution,
openaibaselines}.
This approach imposes a massive systems engineering burden on the development of
distributed applications by essentially pushing standard systems challenges like
scheduling, fault tolerance, and data movement onto each application.

In this paper, we propose {\System}, a general-purpose cluster-computing framework that enables
simulation, training, and serving for RL applications. The requirements of these
workloads range from lightweight and stateless computations, such as for
simulation, to long-running and stateful computations, such as for training.
To satisfy these requirements, {\System} implements a unified interface that can
express both \emph{task-parallel} and \emph{actor-based} computations. \emph{Tasks} enable {\System} to efficiently
and dynamically load balance simulations, process
large inputs and state spaces (e.g., images, video), and recover from failures.  In contrast, \emph{actors} enable {\System} to
efficiently support stateful computations, such as model
training, and expose shared mutable state to clients, (e.g., a parameter
server).
{\System} implements the actor and the task abstractions on top of a
single dynamic execution engine that is highly scalable and fault
tolerant.

To meet the performance requirements, {\System} distributes two components that
are typically centralized in existing
frameworks~\citep{spark-nsdi12,dryad,ciel}: (1) the task scheduler and (2) a
metadata store which maintains the computation lineage and a directory for data
objects.
This allows {\System} to schedule millions of tasks per second with  millisecond-level latencies. Furthermore, {\System} provides lineage-based fault tolerance for tasks and actors, and replication-based fault tolerance for the metadata store.

While {\System} supports serving, training, and simulation in the context
of RL applications,
this does not mean that it should
be viewed as a replacement for systems that provide solutions for these workloads
in other contexts. In particular, {\System} does not aim to substitute for
serving systems like Clipper~\cite{clipper} and TensorFlow
Serving~\cite{tensorflow-serving}, as
these systems address a broader set of challenges in deploying models,
including model management, testing,  and model composition.  Similarly,
despite its flexibility, {\System} is not a substitute for generic
data-parallel frameworks, such as Spark~\cite{spark-nsdi12}, as it
currently lacks the rich functionality and APIs (e.g., straggler
mitigation, query optimization) that these frameworks provide.

We make the following {\bf contributions}:

\begin{itemize}

\item We design and build the first distributed framework that unifies training,
simulation, and serving---necessary components of emerging RL applications.

\item To support these workloads, we unify the actor and task-parallel
abstractions on top of a dynamic task execution engine.

\item To achieve scalability and fault tolerance, we propose a system design
principle in which control state is stored in a sharded metadata store and
all other system components are stateless.

\item To achieve scalability, we propose a bottom-up distributed scheduling
strategy.

\end{itemize}

\begin{figure}
\includegraphics[width=80mm]{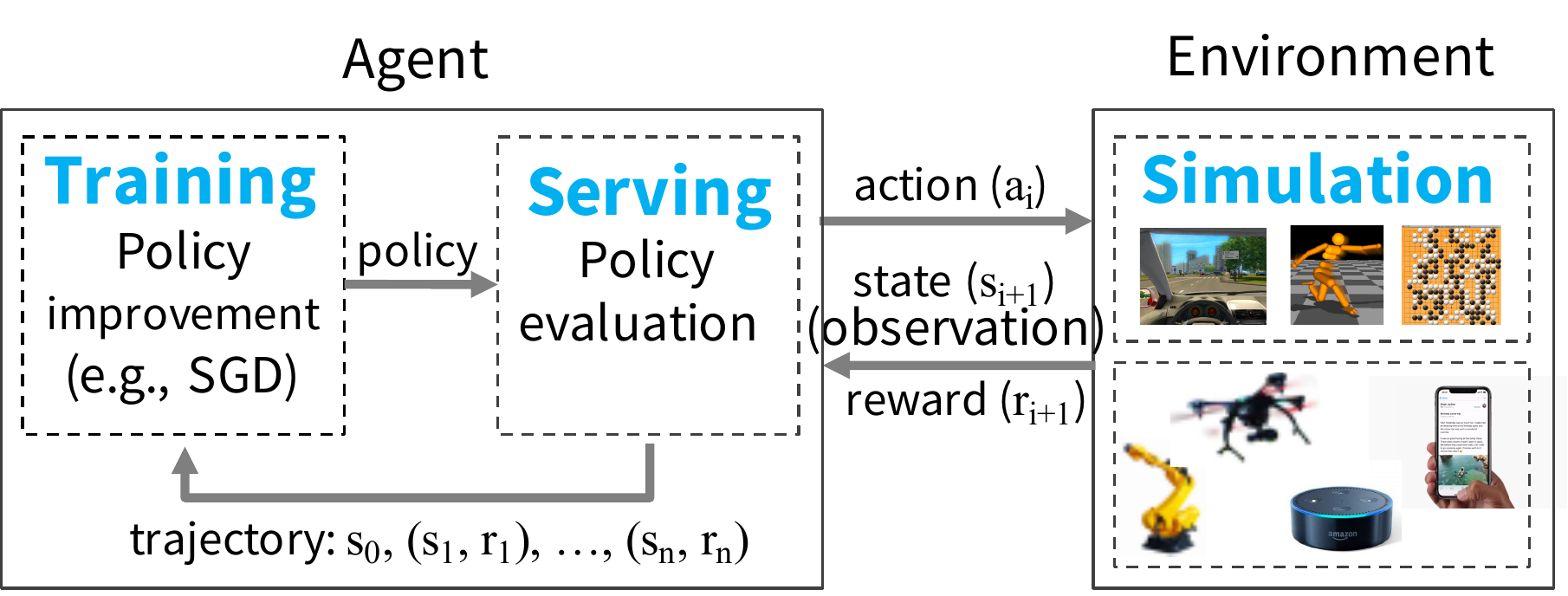}
\caption{\small{Example of an RL system.}}
\label{fig:rl-workloads}
\end{figure}

\section{Motivation and Requirements}
\label{sec:motivation}

We begin by considering the basic components of an RL system and fleshing
out the key requirements for {\System}. As shown in \figref{fig:rl-workloads},
in an RL setting, an \emph{agent}  interacts repeatedly with the
\emph{environment}. The goal of the agent is to learn a policy that maximizes
a \emph{reward}.  A \emph{policy} is a mapping from the state of the
environment  to a choice of \emph{action}. The precise definitions of
environment, agent, state, action, and reward are application-specific.

\begin{figure}
\includegraphics[width=80mm]{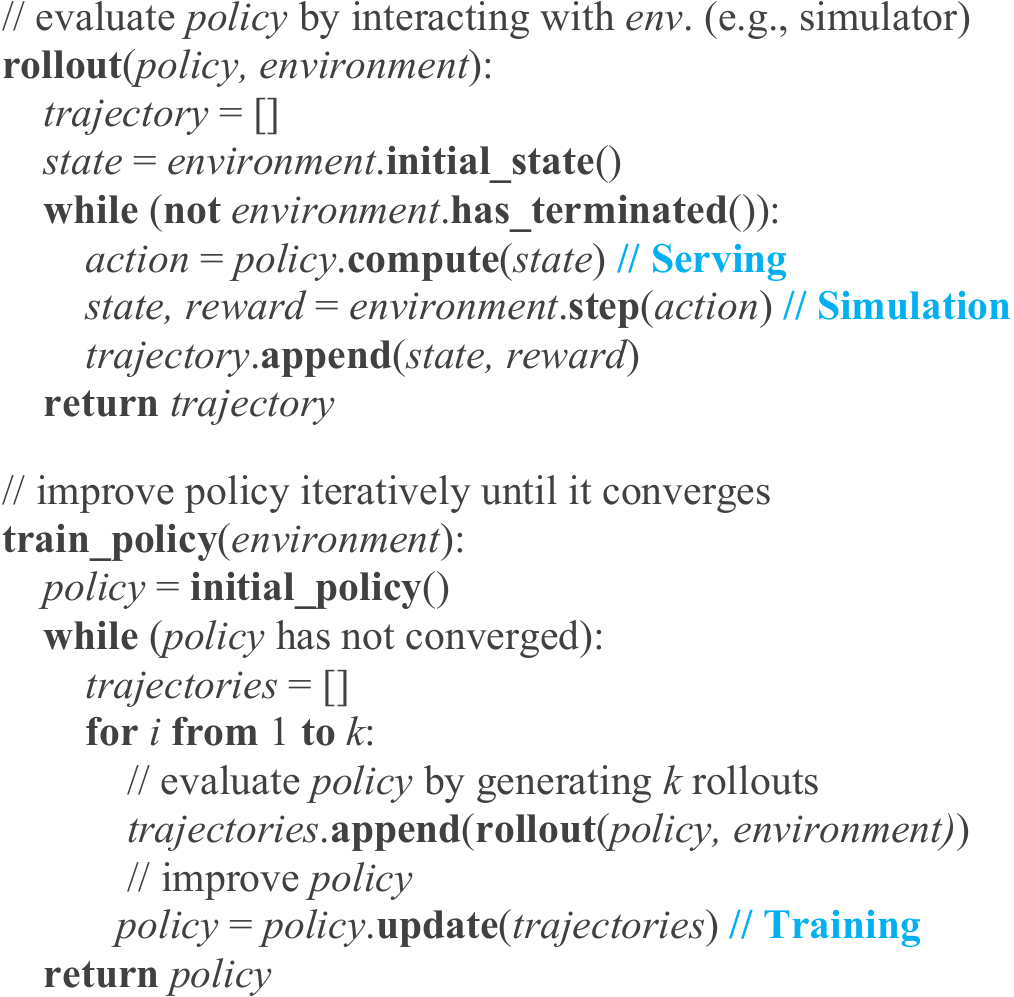}
\caption{\small{Typical RL pseudocode for learning a policy.}}
\label{fig:rl-example-pseudocode}
\end{figure}

To learn a policy, an agent typically employs a two-step process:
(1) \emph{policy evaluation} and (2) {\em policy improvement}. To evaluate the
policy, the agent interacts with the environment (e.g., with a simulation of the
environment) to generate \emph{trajectories}, where a trajectory consists of a sequence of
(state, reward) tuples produced by the current policy. Then, the
agent uses these trajectories to improve the policy; i.e., to update the
policy in the direction of the gradient that maximizes the
reward. \figref{fig:rl-example-pseudocode} shows an example of the
pseudocode used by an agent to learn a policy. This pseudocode
evaluates the policy by invoking {\bf rollout}({\em environment},
{\em policy}) to generate trajectories.
${\bf train\_policy()}$ then uses these trajectories to improve the current
policy via {\em policy}.{\bf update}({\em trajectories}). This process
repeats until the policy converges.

Thus, a framework for RL applications must provide
efficient support for {\em training}, {\em serving}, and
\emph{simulation} (\figref{fig:rl-workloads}). Next, we briefly
describe these workloads.

\emph{Training} typically involves running stochastic gradient descent
(SGD), often in a distributed setting, to update the policy.
Distributed SGD typically relies on an allreduce aggregation step or a
parameter server~\cite{param-server}.

\emph{Serving} uses the trained policy to render an action based on the
current state of the environment. A serving system aims to minimize
latency, and maximize the number of decisions per second.
To scale, load is typically balanced across multiple nodes serving the
policy.

Finally, most existing RL applications use \emph{simulations} to evaluate the
policy---current RL algorithms are not sample-efficient enough to rely solely on
data obtained from interactions with the physical world.
These simulations vary widely in complexity. They might take
a few ms (e.g., simulate a move in a chess game) to
minutes (e.g., simulate a realistic environment for a self-driving
car).

In contrast with supervised learning, in which training and serving can be
handled separately by different systems, in RL \emph{all three of these workloads
are tightly coupled in a single application}, with stringent latency requirements
between them. Currently, no framework supports this coupling of workloads.
In theory, multiple specialized frameworks could be stitched together to
provide the overall capabilities, but in practice, the resulting data
movement and latency between systems is prohibitive in the context of RL.
As a result, researchers and practitioners have been building their own one-off
systems.

This state of affairs calls for the development of new distributed
frameworks for RL that can efficiently support training, serving, and
simulation.
In particular, such a framework should
satisfy the following requirements:

{\it Fine-grained, heterogeneous computations.} The duration of a computation
  can range from milliseconds (e.g., taking an action) to hours (e.g., training
  a complex policy). Additionally, training often requires
  heterogeneous hardware (e.g., CPUs, GPUs, or TPUs).

{\it Flexible computation model.} RL applications require both
  stateless and stateful computations. Stateless computations can be
  executed on any node in the system, which makes it easy to achieve
  load balancing and movement of computation to data, if needed. Thus
  stateless computations are a good fit for fine-grained simulation
  and data processing, such as extracting features from images
  or videos. In contrast stateful computations are a good fit for
  implementing parameter servers, performing repeated computation on
  GPU-backed data, or running third-party simulators that do not
  expose their state.

{\it Dynamic execution.} Several components of RL
  applications require dynamic execution, as the order in which computations
  finish is not always known in advance (e.g., the order in which
  simulations finish), and the results of a computation can determine
  future computations (e.g., the results of a simulation will determine
  whether we need to perform more simulations).

\begin{table*}[tb]
\begin{center}
{\small
\begin{tabular}{| l | l |}
\hline
{\bf Name} & {\bf Description} \\\hline
${\it futures} = {\bf f.remote}({\it args})$ & Execute function $f$ remotely. ${\bf f.remote()}$ can take objects or futures as inputs\\
 & and returns one or more futures. This is non-blocking.\\\hline
${\it objects} = {\bf \system.get}({\it futures})$ & Return the values associated with one or more futures. This is blocking.\\\hline
${\it ready\_futures} = {\bf \system.wait({\it futures}, {\it k}, {\it timeout})}$ & Return the futures whose corresponding tasks have completed as soon as either\\
& $k$ have completed or the timeout expires.\\\hline
${\it actor} = {\bf Class.remote({\it args})}$ & Instantiate class $Class$ as a remote actor, and return a handle to it. Call a method\\
${\it futures} = {\it actor}{\bf .method.remote({\it args})}$ & on the remote actor and return one or more futures. Both are non-blocking.\\\hline
\end{tabular}
}
\end{center}
\caption{{\System} API}
\label{table:api}
\end{table*}

\begin{table}[b]
	\begin{center}
		\begin{footnotesize}
			\begin{tabular}{| c | c |}\hline
				{\bf Tasks (stateless)} & {\bf Actors (stateful)} \\\hline
				Fine-grained load balancing & Coarse-grained load balancing \\\hline
				Support for object locality & Poor locality support \\\hline
				High overhead for small updates & Low overhead for small updates \\\hline
				Efficient failure handling & Overhead from checkpointing \\\hline
			\end{tabular}
		\end{footnotesize}
	\end{center}
	\caption{\small{Tasks vs. actors tradeoffs.}}
	\label{table:tasks-vs-actors}
\end{table}

  We make two final comments. First, to
  achieve high utilization in large clusters,
  such a framework must handle {\em millions of tasks per second}.\footnote{Assume 5ms
    single-core tasks and a cluster of 200 32-core
    nodes. This cluster can run
    $(1s/5ms) \times 32 \times 200 = 1.28$M tasks/sec.}
  Second, such a framework is not intended for
  implementing deep neural networks or complex simulators from
  scratch.
  Instead, it should enable seamless integration with existing
  simulators \cite{brockman2016openai, beattie2016deepmind, todorov2012mujoco}
  and deep learning frameworks \cite{tensorflow-osdi16,
  mxnet-learningsys, pytorch, jia2014caffe}.

\section{Programming and Computation Model}
\label{sec:design}

\begin{figure}[tb]
	\centering
		\centering
		\begin{footnotesize}
			\input{fig/rl_code_example.tex}
		\end{footnotesize}
		\caption{
			{\small
				Python code implementing the example in \figref{fig:rl-example-pseudocode} in {\System}. Note that {@\system.remote} indicates remote functions and actors. Invocations of remote functions and actor methods return futures, which can be passed to subsequent remote functions or actor methods to encode task dependencies. Each actor has an environment object {self.env} shared between all of its methods.
			}
		}
		\label{fig:design:rl:code}
		\label{fig:training-python-example}
\end{figure}
\begin{figure}[tb]
		\centering
		\includegraphics[width=0.85\columnwidth]{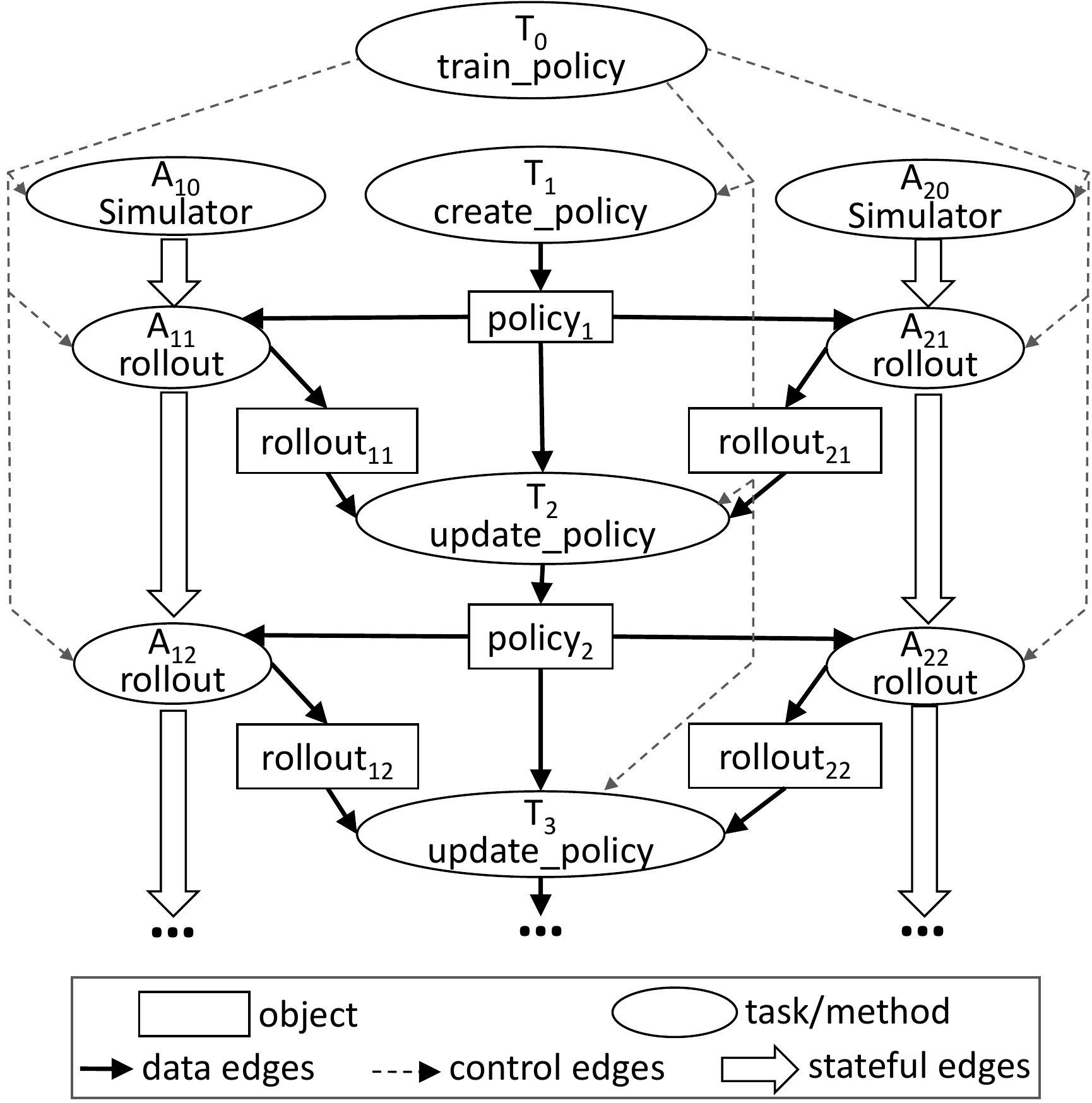}
		\caption{
			{ \small
				 The task graph corresponding to an invocation of {train\_policy.remote()} in \figref{fig:design:rl:code}. Remote function calls and the actor method calls correspond to tasks in the task graph. The figure shows two actors. The method invocations for each actor (the tasks labeled $A_{1i}$ and $A_{2i}$) have stateful edges between them indicating that they share the mutable actor state. There are control edges from {train\_policy} to the tasks that it invokes. To train multiple policies in parallel, we could call {train\_policy.remote()} multiple times.
			}
		}
		\label{fig:design:rl:dataflow}
\end{figure}

{\System} implements a dynamic task graph computation model, i.e.,
it models an application as a graph of dependent tasks that evolves during execution. On top of this
model, {\System} provides both an actor and a task-parallel programming
abstraction. This unification differentiates {\System} from related
systems like CIEL, which only provides a task-parallel abstraction, and from
Orleans~\cite{bykov2011orleans} or Akka~\cite{akka}, which primarily provide an
actor abstraction.

 \subsection{Programming Model}
\label{sec:api}

\noindent {\bf Tasks.} A \emph{task} represents the execution of a remote
function on a stateless worker. When a remote function is invoked, a \emph{future}
representing the result of the task is returned immediately. Futures can be
retrieved using ${\bf \system.get()}$ and passed as arguments into other remote
functions without waiting for their result. This allows the user to express
parallelism while capturing data dependencies. \tabref{table:api} shows
{\System}'s API.

Remote functions operate on immutable objects and are expected to be
\emph{stateless} and side-effect free: their outputs are determined solely by
their inputs. This implies idempotence, which simplifies fault tolerance through
function re-execution on failure.

\noindent {\bf Actors.} An \emph{actor} represents a stateful computation. Each actor exposes methods that
can be invoked remotely and are executed serially.
A method execution is similar to a task, in that it executes remotely and returns a future, but differs in that it executes on
a {\em stateful} worker. A {\em handle} to an actor can be passed to other actors or
tasks, making it possible for them to invoke methods on that actor.

\tabref{table:tasks-vs-actors}  summarizes the properties of tasks and actors. Tasks enable fine-grained load balancing through
leveraging load-aware scheduling at task granularity, input data locality, as each task can be scheduled on the node storing its inputs, and low recovery overhead, as there is no need to checkpoint and recover intermediate state. In contrast, actors provide much more efficient fine-grained updates, as these updates are performed on internal rather than external state, which typically requires serialization and deserialization. For example, actors can be used to implement parameter servers~\citep{param-server} and GPU-based iterative computations (e.g., training). In addition, actors can be used to wrap third-party simulators and other opaque handles that are hard to serialize.

To satisfy the requirements for heterogeneity and flexibility (\secref{sec:motivation}), we augment the API in three ways. First, to handle concurrent tasks with heterogeneous durations, we introduce ${\bf \system.wait()}$, which waits for the first $k$ available results, instead of waiting for {\em all} results like ${\bf \system.get()}$.
Second, to handle resource-heterogeneous tasks, we enable developers to specify resource requirements so that the {\System} scheduler can efficiently manage resources.
Third, to improve flexibility, we enable {\em nested remote functions}, meaning that remote functions can invoke other remote functions. This is also critical for achieving high scalability (\secref{sec:arch}), as it enables multiple processes to invoke remote functions in a distributed fashion.

\subsection{Computation Model}

{\System} employs a dynamic task graph computation model~\cite{dataflow74}, in which the execution of both remote functions and actor methods is automatically triggered by the system when their inputs become available. In this section, we describe how the computation graph (\figref{fig:design:rl:dataflow}) is constructed from a user program (\figref{fig:training-python-example}). This program uses the API in \tabref{table:api} to implement the pseudocode from \figref{fig:rl-example-pseudocode}.

Ignoring actors first, there are two types of nodes in a computation graph: data
objects and remote function invocations, or tasks. There are also two types of
edges: data edges and control edges. Data edges capture the dependencies
between data objects and tasks. More precisely, if data object $D$ is an output of task $T$, we
add a data edge from $T$ to $D$. Similarly, if $D$ is an input to
$T$, we add a data edge from $D$ to $T$.
Control edges capture the computation dependencies that result from nested
remote functions (\secref{sec:api}): if task $T_1$ invokes task $T_2$, then we
add a control edge from $T_1$ to $T_2$.

Actor method invocations are also represented as nodes in the computation graph. They are identical to tasks with one key difference. To capture the state dependency across subsequent method invocations on the same actor, we add a third type of edge: a stateful edge. If method $M_j$ is called right after method $M_i$ on the same actor, then  we add a stateful edge from $M_i$ to $M_j$. Thus, all methods invoked on the same actor object form a chain that is connected by stateful edges (\figref{fig:design:rl:dataflow}). This chain captures the order in which these methods were invoked.

Stateful edges help us embed actors in an otherwise stateless task graph, as
they capture the implicit data dependency between successive method invocations sharing
the internal state of an actor. Stateful edges also enable us to maintain
lineage. As in other dataflow systems~\citep{spark-nsdi12}, we track data
lineage to enable reconstruction. By explicitly including stateful edges in the
lineage graph, we can easily reconstruct lost data, whether produced by remote
functions or actor methods (\secref{sec:object-store}).

\section{Architecture}

{\System}'s architecture comprises (1) an application layer
implementing the API, and (2) a system layer providing high scalability
and fault tolerance.

\label{sec:arch}

\begin{figure}[h]
\includegraphics[width=80mm]{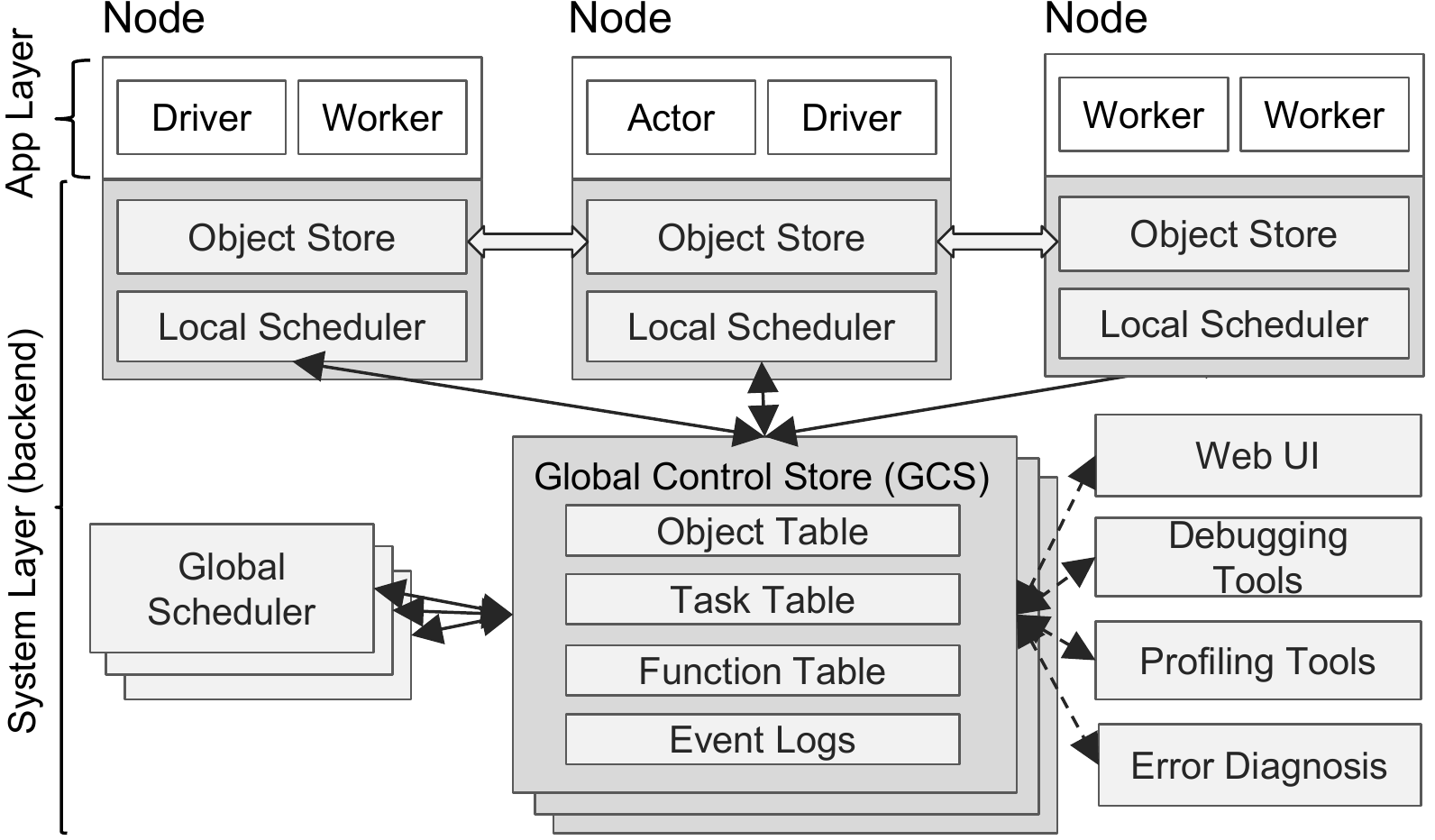}
\caption{\small{{\System}'s architecture consists of two parts:
an {\em application} layer and a {\em system} layer. The
application layer implements the API and the computation model described in
\secref{sec:design}, the system layer implements task scheduling and data
management to satisfy the performance and fault-tolerance requirements.}}
\label{fig:ray-architecture}
\end{figure}

\subsection{Application Layer}
\label{sec:app-layer}

The application layer consists of three types of processes:
\begin{itemize}
\item {\em Driver}: A process executing the user program.
\item {\em Worker}: A stateless process that executes tasks (remote functions)
invoked by a driver or another worker. Workers are started automatically and
assigned tasks by the system layer. When a remote function is declared, the
function is automatically published to all workers. A worker executes tasks
serially, with no local state maintained across tasks.
\item {\em Actor}: A stateful process that executes, when invoked, only the methods
it exposes. Unlike a worker, an actor is explicitly instantiated by a worker or
a driver. Like workers, actors execute methods serially, except that each
method depends on the state resulting from the previous method execution.
\end{itemize}

\subsection{System Layer}
\label{sec:sys-layer}
The system layer consists of three major components: a global control store, a distributed scheduler, and
a distributed object store. All components are horizontally scalable and fault-tolerant.

\subsubsection{Global Control Store (GCS)}
\label{sec:gcs}
The global control store (GCS) maintains the entire control state of the system, and it is a unique feature of our design.
At its core, GCS is a key-value store with pub-sub functionality.
We use sharding to achieve scale, and per-shard chain replication~\cite{chain-replication} to provide fault tolerance.
The primary reason for the GCS and its design is to maintain fault tolerance and low latency for a system that can dynamically spawn millions of tasks per second.

Fault tolerance in case of node failure requires a solution to maintain lineage information.
Existing lineage-based solutions~\citep{spark-nsdi12,hadoop,ciel,dryad} focus on coarse-grained parallelism and can therefore use a single node (e.g., master, driver) to store the lineage without impacting performance.
However, this design is not scalable for a fine-grained and dynamic workload like simulation.
Therefore, we decouple the durable lineage storage from the other system components, allowing each to scale independently.

Maintaining low latency requires minimizing overheads in task scheduling, which involves choosing where to execute, and subsequently task dispatch, which involves retrieving remote inputs from other nodes.
Many existing dataflow systems~\citep{spark-nsdi12,ciel,dask-scipy15} couple these by storing object locations and sizes in a centralized scheduler, a natural design when the scheduler is not a bottleneck.
However, the scale and granularity that \System{} targets requires keeping the centralized scheduler off the critical path.
Involving the scheduler in each object transfer is prohibitively expensive for primitives important to distributed training like allreduce, which is both communication-intensive and latency-sensitive.
Therefore, we store the object metadata in the GCS rather than in the scheduler, fully decoupling task dispatch from task scheduling.

In summary, the GCS significantly simplifies \System{}'s overall design, as it~\emph{enables every component in the system to be stateless}.
This not only simplifies support for fault tolerance (i.e., on failure, components simply restart and read the lineage from the GCS), but also makes it easy to scale the distributed object store and scheduler independently, as all components share the needed state via the GCS.
An added benefit is the easy development of debugging, profiling, and visualization tools.

\subsubsection{Bottom-Up Distributed Scheduler}
\label{sec:arch:sched}
As discussed in Section~\ref{sec:motivation}, {\System} needs to dynamically schedule millions of tasks per second, tasks which may take as little as a few milliseconds.
None of the cluster schedulers we are aware of meet these requirements. Most cluster computing frameworks, such as  Spark~\citep{spark-nsdi12}, CIEL~\citep{ciel}, and Dryad~\citep{dryad} implement a centralized scheduler, which can provide locality but at latencies in the tens of ms. Distributed schedulers such as work stealing~\cite{work-stealing}, Sparrow~\citep{sparrow} and Canary~\citep{qu2016canary} can achieve high scale, but they either don't consider data locality~\cite{work-stealing}, or assume tasks belong to independent jobs~\citep{sparrow}, or assume the computation graph is known~\citep{qu2016canary}.

\begin{figure}[h]
\includegraphics[width=75mm]{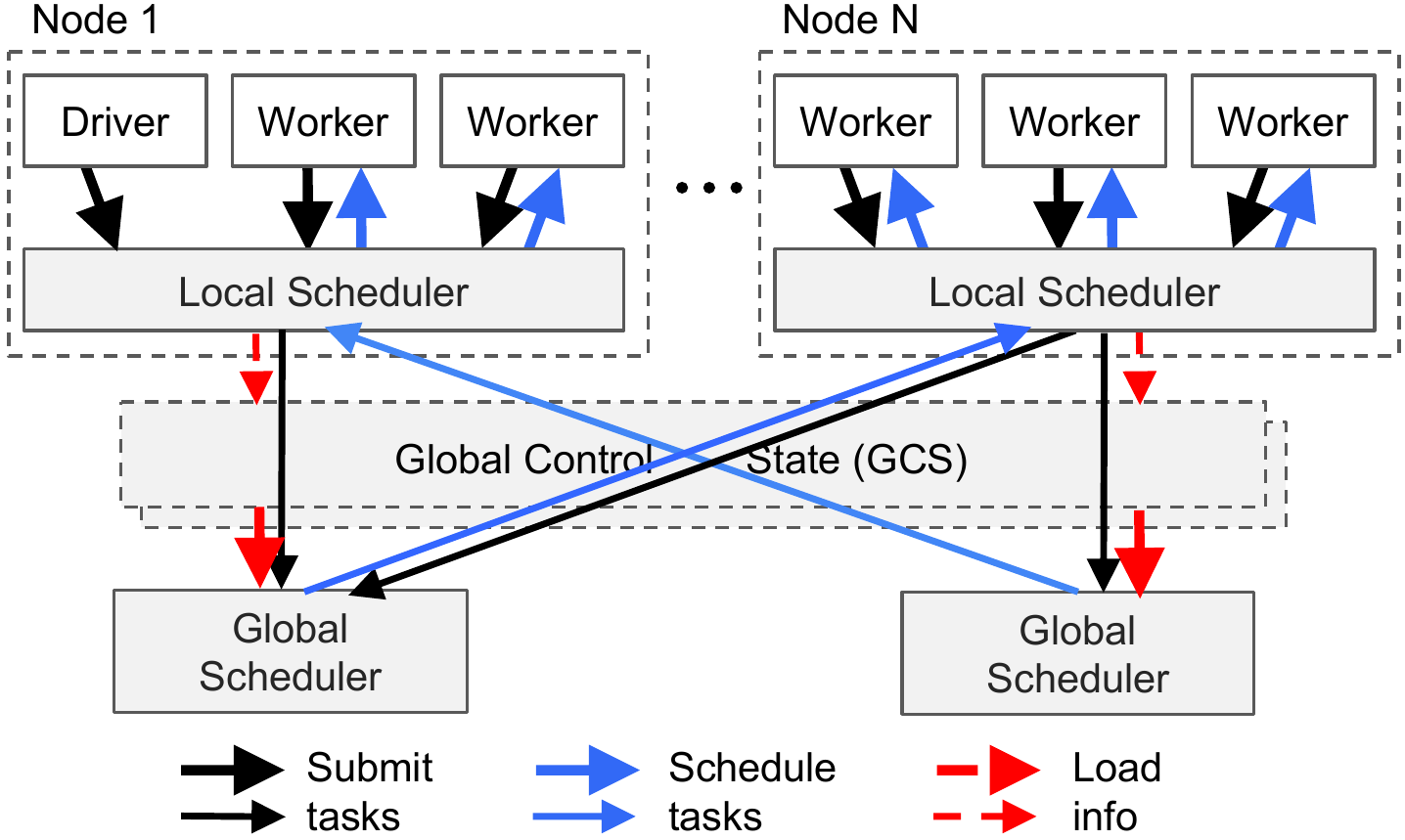}
\caption{\small{Bottom-up distributed scheduler.
Tasks are submitted bottom-up, from drivers and workers to a local scheduler and
forwarded to the global scheduler only if needed (\secref{sec:arch:sched}).
The thickness of each arrow is proportional to its request rate.}}
\label{fig:scheduler}
\end{figure}

To satisfy the above requirements, we design a two-level hierarchical scheduler consisting of a global scheduler and per-node local schedulers.
To avoid overloading the global scheduler, the tasks created at a node are submitted first to the node's local scheduler.
A local scheduler schedules tasks locally unless the node is overloaded (i.e., its local task queue exceeds a predefined threshold), or it cannot satisfy a task's requirements (e.g., lacks a GPU).
If a local scheduler decides not to schedule a task locally, it forwards it to the global scheduler.
Since this scheduler attempts to schedule tasks locally first (i.e., at the leaves of the scheduling hierarchy), we call it a {\em bottom-up scheduler}.

\looseness=-1
The global scheduler considers each node's load and task's constraints to make scheduling decisions. More precisely, the global scheduler identifies the set of nodes that have enough resources of the type requested by the task, and of these nodes selects the node which provides the lowest \emph{estimated waiting time}. At a given node, this time is the sum of (i) the estimated time the task will be queued at that node (i.e., task queue size times average task execution), and (ii) the estimated transfer time of task’s remote inputs (i.e., total size of remote inputs divided by average bandwidth). The global scheduler gets the queue size at each node and the node resource availability via heartbeats, and the location of the  task's inputs and their sizes from GCS. Furthermore, the global scheduler computes the average task execution and the average transfer bandwidth using simple exponential averaging.
If the global scheduler becomes a bottleneck, we can instantiate more replicas all sharing the same information via GCS. This makes our scheduler architecture highly scalable.

\subsubsection{In-Memory Distributed Object Store}
\label{sec:object-store}
To minimize task latency, we implement an in-memory distributed storage system to store the inputs and outputs of every task, or stateless computation.
On each node, we implement the object store via {\em shared memory}. This allows zero-copy data sharing between tasks running on the same node. As a data format, we use Apache Arrow~\cite{arrow}.

If a task's inputs are not local, the inputs are replicated to the local object store before execution. Also, a task writes its outputs to the local object store. Replication eliminates the potential bottleneck due to hot data objects and minimizes task execution time as a task only reads/writes data from/to the local memory. This increases throughput for computation-bound workloads, a profile shared by many AI applications.
For low latency, we keep objects entirely in memory and evict them as needed to disk using an LRU policy.

As with existing cluster computing frameworks, such as Spark~\citep{spark-nsdi12}, and Dryad~\citep{dryad}, the object store is limited to {\em immutable data}. This obviates the need for complex consistency protocols (as objects are not updated), and simplifies support for fault tolerance.
In the case of node failure, \System{} recovers any needed objects through lineage re-execution.
The lineage stored in the GCS tracks both stateless tasks and stateful actors during initial execution; we use the former to reconstruct objects in the store.

For simplicity, our object store does not support distributed objects, i.e., each object fits on a single node. Distributed objects like large matrices or trees can be implemented at the application level as collections of futures.

\subsubsection{Implementation}
\System{} is an active open source project\footnote{\url{https://github.com/ray-project/ray}} developed at the University of California, Berkeley.
\System{} fully integrates with the Python environment and is easy to install by simply running \texttt{pip install ray}.
The implementation comprises $\approx~40$K lines of code (LoC), 72\% in C++ for the system layer, 28\% in Python for the application layer.
The GCS uses one Redis \citep{redis2009} key-value store per shard, with entirely single-key operations.
GCS tables are sharded by object and task IDs to scale, and every shard is chain-replicated~\cite{chain-replication} for fault tolerance.
We implement both the local and global schedulers as event-driven, single-threaded processes.
Internally, local schedulers maintain cached state for local object metadata, tasks waiting for inputs, and tasks ready for dispatch to a worker.
To transfer large objects between different object stores, we stripe the object across multiple TCP connections.

\begin{figure}[tb]
    \centering
    \begin{subfigure}[b]{\columnwidth}
        \centering
        \includegraphics[width=\textwidth]{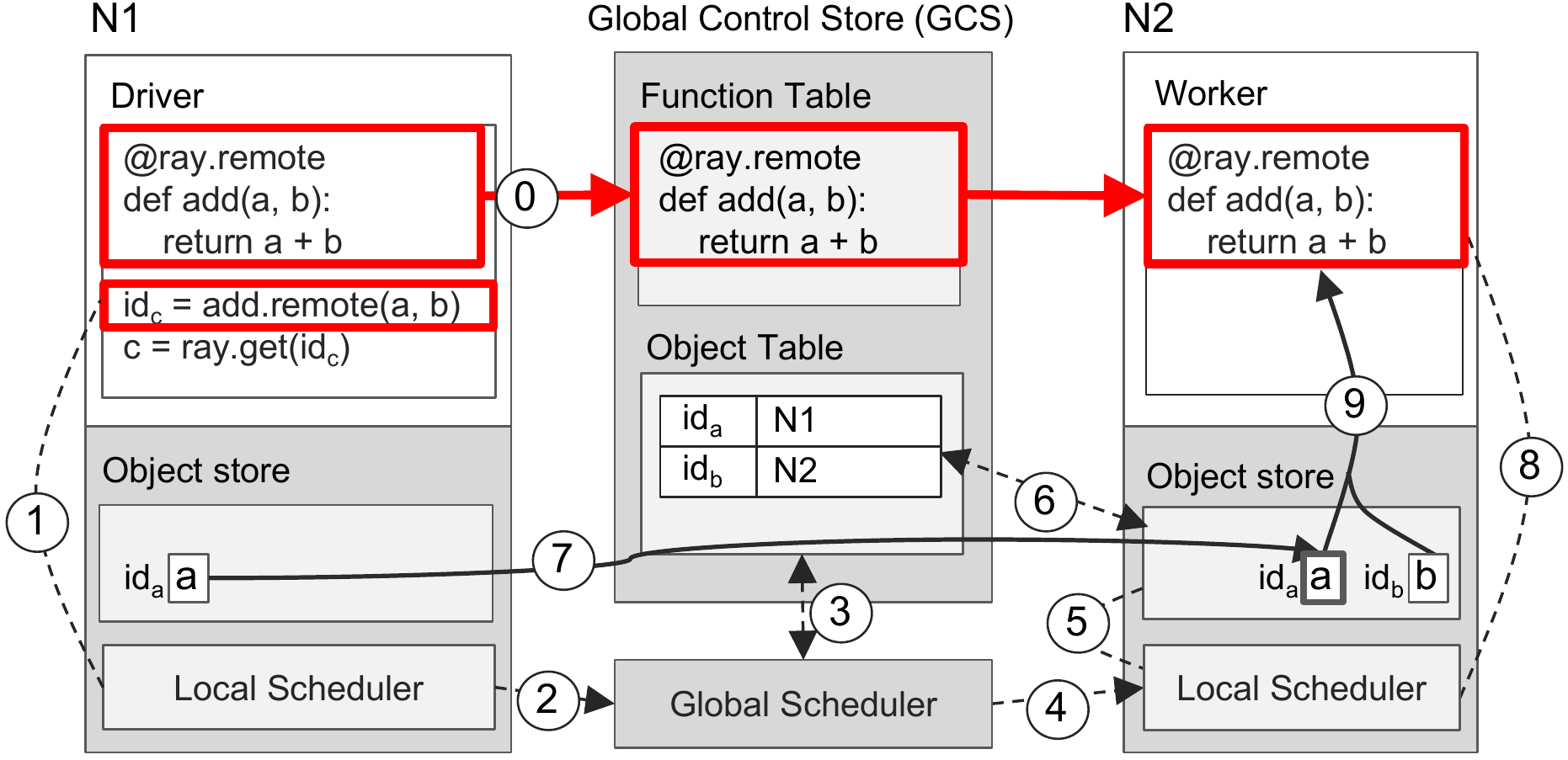}
        \caption{Executing a task remotely}
    \end{subfigure}
    \begin{subfigure}[b]{\columnwidth}
        \centering
        \includegraphics[width=\textwidth]{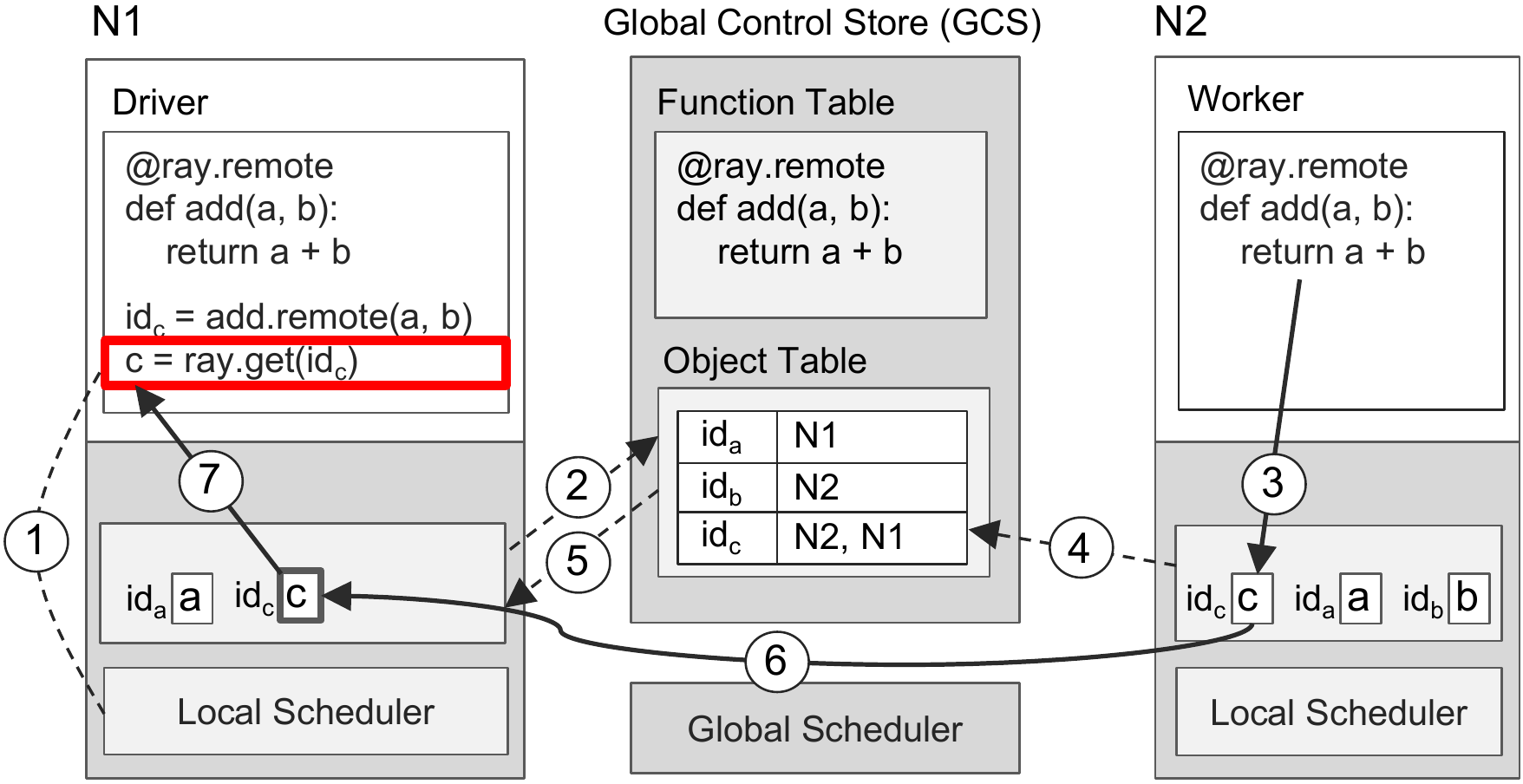}
        \caption{Returning the result of a remote task}
    \end{subfigure}
\caption{\small{An end-to-end example that adds $a$ and $b$ and returns $c$.
Solid lines are data plane operations and dotted lines are control plane operations.
(a) The function {\bf add()} is registered with the GCS by node 1 ($N1$), invoked on $N1$, and executed on
$N2$. (b) $N1$ gets {\bf add()}'s result using
{\bf \system.get()}. The Object Table entry for $c$ is created in step 4
and updated in step 6 after $c$ is copied to $N1$.}}
\label{fig:step-by-step}
\end{figure}

\subsection{Putting Everything Together}
\label{sec:everything}

\figref{fig:step-by-step} illustrates how {\System} works end-to-end with a
simple example that adds two objects $a$ and $b$, which could be scalars or
matrices, and returns result $c$. The remote function {\bf add()} is
automatically registered with the GCS upon initialization and distributed to
every worker in the system (step 0 in \figref{fig:step-by-step}a).

\figref{fig:step-by-step}a shows the step-by-step operations triggered by a
driver invoking {\bf add.remote($a, b$)}, where $a$ and $b$ are stored on nodes
$N1$ and $N2$, respectively. The driver submits {\bf add($a$, $b$)} to the local
scheduler (step 1), which forwards it to a global scheduler
(step 2).\footnote{Note that $N1$ could also decide to schedule
the task locally.} Next, the global scheduler looks up the locations of
{\bf add($a$, $b$)}'s arguments in the GCS (step 3) and decides to schedule the
task on node $N2$, which stores argument $b$ (step 4). The local scheduler at
node $N2$ checks whether the local object store contains {\bf add($a$, $b$)}'s
arguments (step 5). Since the local store doesn't have object $a$, it looks up
$a$'s location in the GCS (step 6). Learning that $a$ is stored at $N1$, $N2$'s
object store replicates it locally (step 7). As all arguments of {\bf add()} are
now stored locally, the local scheduler invokes {\bf add()} at a local worker
(step 8), which accesses the arguments via shared memory (step 9).

\figref{fig:step-by-step}b shows the step-by-step operations triggered by the
execution of {\bf \system.get()} at $N1$, and of {\bf add()} at $N2$,
respectively. Upon {\bf \system.get($id_c$)}'s invocation, the driver checks
the local object store for the value $c$, using the future $id_c$ returned by
{\bf add()} (step 1). Since the local object store doesn't store $c$, it looks
up its location in the GCS. At this time, there is no entry for $c$, as $c$ has
not been created yet. As a result, $N1$'s object store registers a callback
with the Object Table to be triggered when $c$'s entry has been created
(step 2). Meanwhile, at $N2$, {\bf add()} completes its execution, stores the
result $c$ in the local object store (step 3), which in turn adds $c$'s entry
to the GCS (step 4). As a result, the GCS triggers a callback to $N1$'s object
store with $c$'s entry (step 5). Next, $N1$ replicates $c$ from $N2$ (step 6),
and returns $c$ to {\bf \system.get()} (step 7), which finally completes the
task.

While this example involves a large number of RPCs, in many cases this
number is much smaller, as most tasks are scheduled locally, and
the GCS replies are cached by the global and local schedulers.

\section{Evaluation}
\label{sec:experiments}

In our evaluation, we study the following questions:

\begin{enumerate}[noitemsep,topsep=0pt,parsep=0pt,partopsep=0pt]

  \item How well does \System{} meet the latency, scalability, and fault
  tolerance requirements listed in \secref{sec:motivation}?
  (\secref{sec:microbenchmarks})

  \item What overheads are imposed on distributed primitives (e.g.,
  allreduce) written using \System{}'s API? (\secref{sec:microbenchmarks})

  \item In the context of RL workloads, how does \System{} compare against
  specialized systems for training, serving, and simulation? (\secref{sec:building_blocks})

  \item What advantages does \System{} provide for RL
  applications, compared to custom systems? (\secref{sec:exp_rl})

\end{enumerate}

All experiments were run on Amazon Web Services.
Unless otherwise stated, we use m4.16xlarge CPU instances and p3.16xlarge GPU
instances.

\begin{figure}
    \centering
    \begin{subfigure}[b]{0.218\textwidth}
        \centering
        \includegraphics[width=3.9cm]{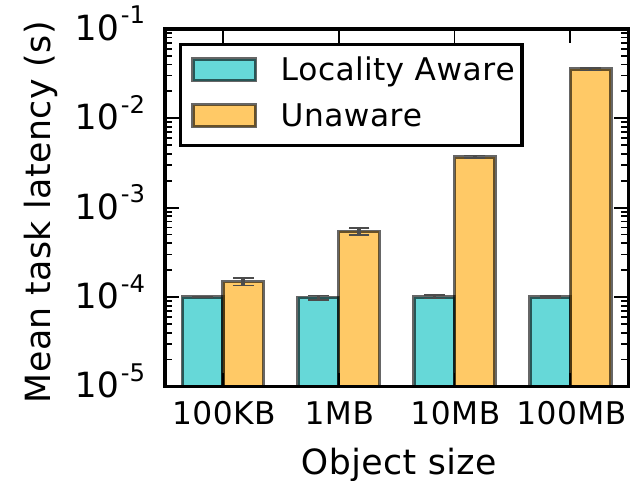}
        \caption{
            \System{} locality scheduling
        }
        \label{fig:sched:locality}
    \end{subfigure}
    \hspace{.4cm}
    \begin{subfigure}[b]{0.218\textwidth}
        \centering
        \includegraphics[width=3.8cm]{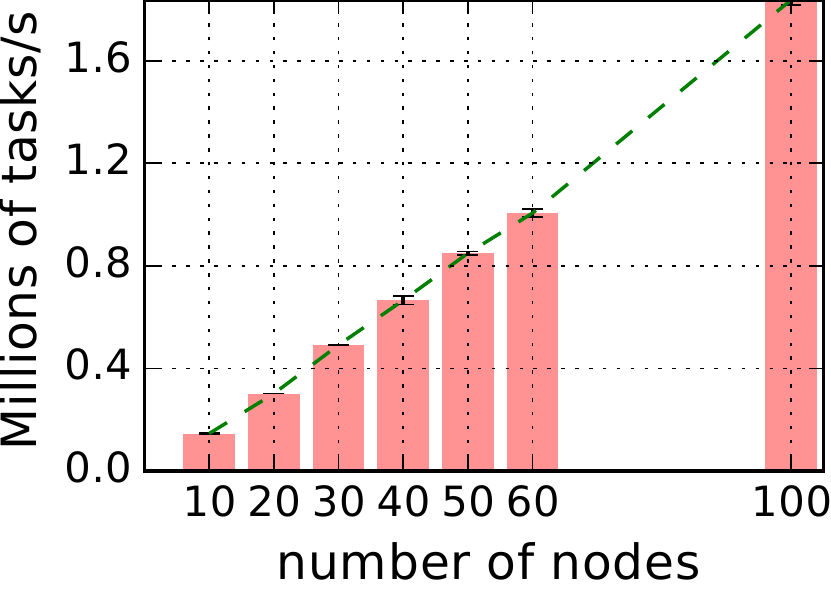}
        \caption{
            \System{} scalability
        }
        \label{fig:sched:scale}
    \end{subfigure}
    \caption{
      \small{
      (a)
        Tasks leverage locality-aware placement.
        1000 tasks with a random object dependency are scheduled onto one of two nodes.
        With locality-aware policy, task latency remains independent of the
        size of task inputs instead of growing by 1-2 orders of magnitude.
       (b) Near-linear scalability leveraging the GCS and bottom-up distributed scheduler.
    \System{} reaches 1 million tasks per second throughput with 60 nodes.
    $x \in \{70, 80, 90\}$ omitted due to cost.
    }
    }
    \label{fig:sched}
\end{figure}

\subsection{Microbenchmarks}
\label{sec:microbenchmarks}

\noindent\hspace{1em}\textbf{Locality-aware task placement.}
Fine-grain load balancing and locality-aware placement are primary benefits of tasks in \System{}.
Actors, once placed, are unable to move their computation to large remote objects, while tasks can.
In \figref{fig:sched:locality}, tasks placed without data locality awareness (as is the case for actor methods), suffer 1-2 orders of magnitude latency increase at 10-100MB input data sizes.
\System{} unifies tasks and actors through the shared object store, allowing developers to use tasks for
e.g., expensive postprocessing on output produced by simulation actors.

\hspace{1em}\textbf{End-to-end scalability.}
One of the key benefits of the Global Control Store (GCS) and the bottom-up distributed scheduler
is the ability to horizontally scale the system to support a high throughput of fine-grained tasks,
while maintaining fault tolerance and low-latency task scheduling.
In Figure~\ref{fig:sched:scale}, we evaluate this ability on an embarrassingly parallel
workload of empty tasks, increasing the cluster size on the x-axis. We observe near-perfect linearity in progressively increasing
task throughput. \System{} exceeds 1 million tasks per second throughput at 60 nodes and continues to scale linearly
beyond 1.8 million tasks per second at 100 nodes. The rightmost datapoint shows that \System{} can process 100 million
tasks in less than a minute (54s), with minimum variability.
As expected, increasing task duration reduces throughput proportionally to mean task duration,
but the overall scalability remains linear.
While many realistic workloads may exhibit more limited scalability due to
object dependencies and inherent limits to application parallelism, this demonstrates
the scalability of our overall architecture under high load.

\begin{figure}[tb]
	\centering
	\includegraphics[width=\columnwidth]{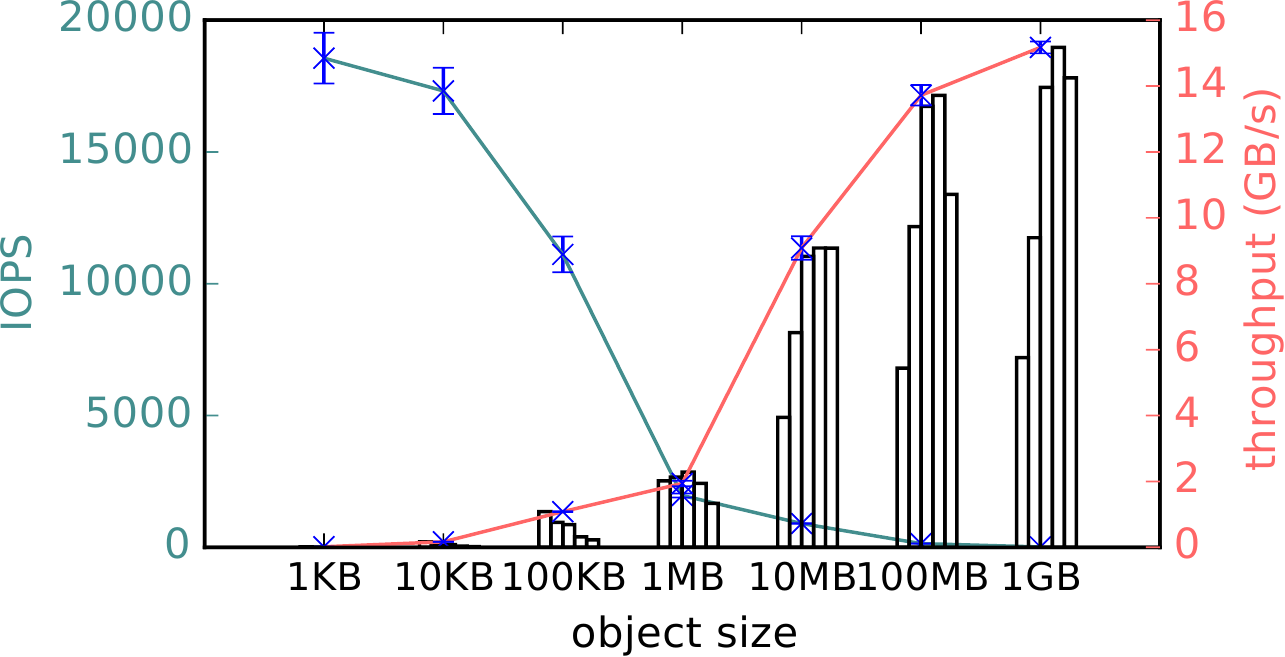}
	\caption{ \small{
			Object store write throughput and IOPS.
			From a single client, throughput exceeds 15GB/s (red) for large
			objects and 18K IOPS (cyan) for small objects on a 16 core instance
			(m4.4xlarge). It uses 8 threads to copy objects larger than 0.5MB and 1 thread for
			small objects. Bar plots report throughput with 1, 2, 4, 8, 16 threads.
			Results are averaged over 5 runs.
		}
	}
	\label{fig:eval:object_store}
\end{figure}

\textbf{Object store performance.}
To evaluate the performance of the object store~(\secref{sec:object-store}),
we track two metrics: IOPS (for small objects) and write throughput (for large objects).
In Figure~\ref{fig:eval:object_store},
the write throughput from a single client exceeds 15GB/s as object size increases.
For larger objects, memcpy dominates object creation time. For smaller objects,
the main overheads are in serialization and IPC between the client and object store.

\begin{figure}[tb]
	\centering
		\begin{subfigure}[b]{\columnwidth}
			\centering
			\includegraphics[width=1.0\columnwidth]{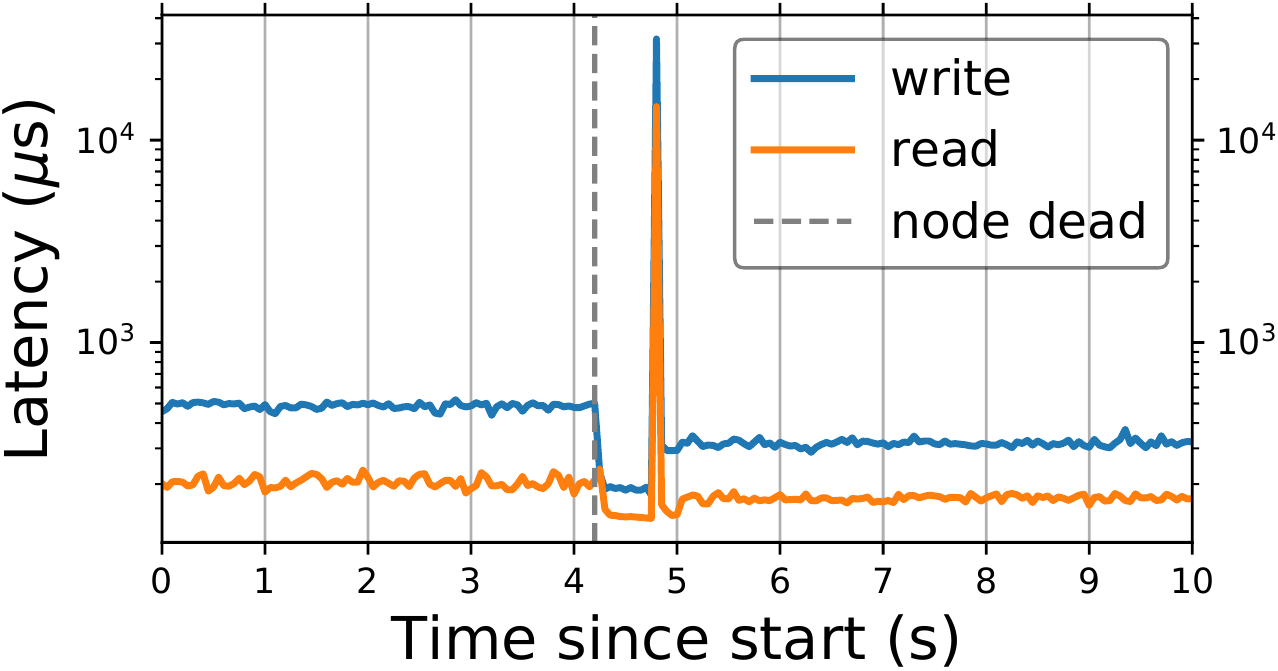}
			\caption{
				{\small A timeline for GCS read and write latencies as viewed from a client submitting tasks.
					The chain starts with 2 replicas.  We manually trigger reconfiguration
					as follows. At $t \approx 4.2$s, a chain member is killed; immediately
					after, a new chain member joins, initiates state transfer, and restores
					the chain to 2-way replication. The
					maximum client-observed latency is under 30ms despite reconfigurations.
				}
			}
			\label{fig:chain_fault_tolerance}
		\end{subfigure}
		\begin{subfigure}[b]{\columnwidth}
			\centering
			\includegraphics[width=1.0\columnwidth]{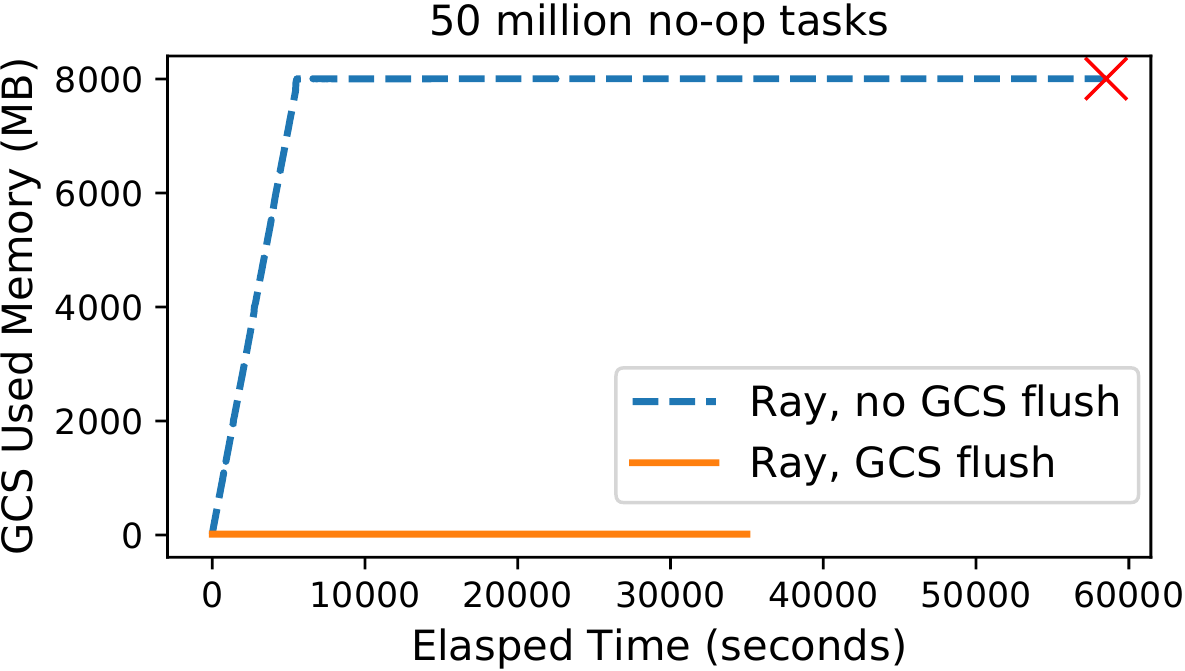}
			\caption{
				{\small The Ray GCS maintains a constant memory footprint with GCS flushing.
					Without GCS flushing, the memory footprint reaches a maximum capacity
          and the workload fails to complete within a predetermined duration
          (indicated by the red cross).
				}
			}
			\label{fig:eval:gcs-flushing}
		\end{subfigure}
	\caption{\small{ \System{} GCS fault tolerance and flushing.}}
	\label{fig:eval:gcs}
\end{figure}

\begin{figure}[tbh]
	\centering
	\begin{subfigure}[b]{\columnwidth}
		\centering
		\includegraphics[width=1.0\columnwidth]{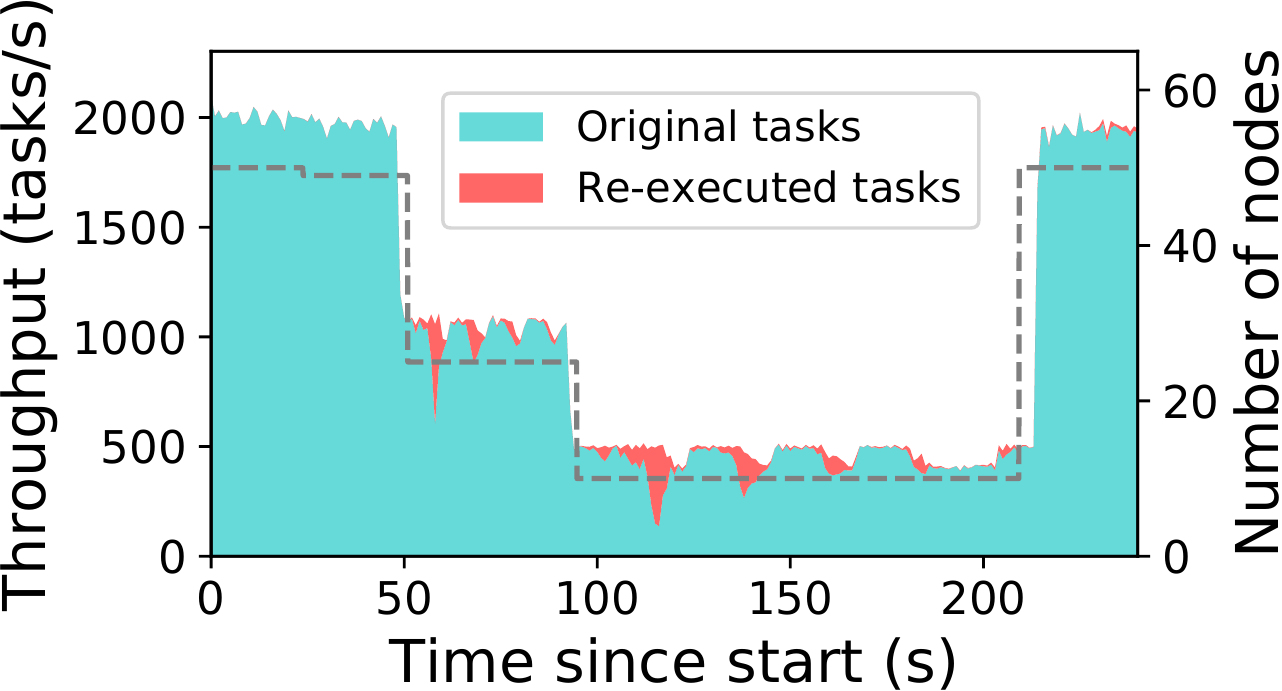}
		\caption{
			Task reconstruction
		}
		\label{fig:fault_tolerance}
	\end{subfigure}
	\begin{subfigure}[b]{\columnwidth}
		\centering
		\includegraphics[width=1.0\columnwidth]{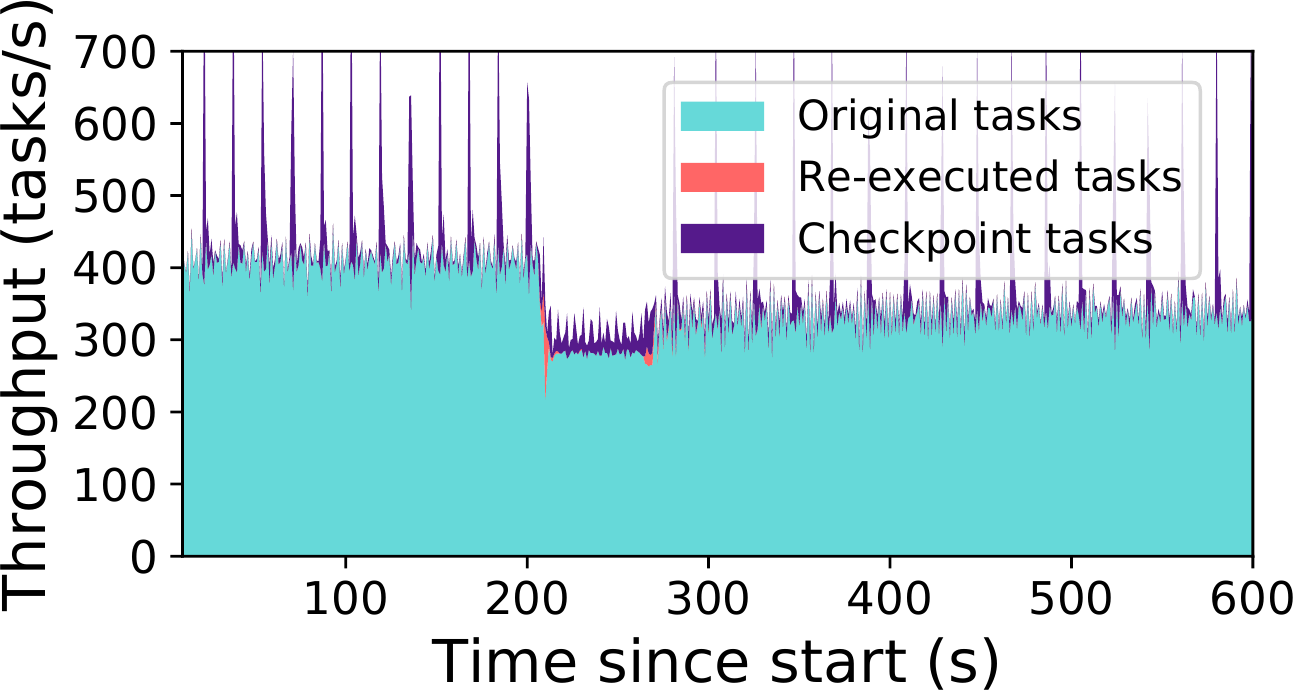}
		\caption{
			Actor reconstruction
		}
		\label{fig:actor_fault_tolerance}
	\end{subfigure}
	    \caption{
		\small{ \System{} fault-tolerance.
			\textbf{(a)} \System{} reconstructs lost task dependencies as nodes are removed
			(dotted line), and recovers to original throughput
			when nodes are added back.
			Each task is 100ms and depends on an object
			generated by a previously submitted task.
			\textbf{(b)} Actors are reconstructed from their last checkpoint. At $t={}$200s,
			we kill 2 of the 10 nodes, causing 400 of the 2000 actors in the cluster to
			be recovered on the remaining nodes ($t={}$200--270s).
		}
	}
	\label{fig:eval:reconstruction}
\end{figure}

\textbf{GCS fault tolerance.}
To maintain low latency while providing strong
consistency and fault tolerance, we build a lightweight chain
replication~\cite{chain-replication} layer on top of Redis.
\figref{fig:chain_fault_tolerance} simulates recording \System{} tasks to
and reading tasks from the GCS, where keys are 25 bytes and values are 512
bytes. The client sends requests as fast as it can, having at most one in-flight
request at a time.  Failures are reported to the chain master either from the
client (having received explicit errors, or timeouts despite retries) or from
any server in the chain (having received explicit errors).
Overall, reconfigurations caused a maximum \emph{client-observed} delay of under
30ms (this includes both failure detection and recovery delays).

\textbf{GCS flushing.}  Ray is equipped to periodically flush the contents of
GCS to disk. In Figure~\ref{fig:eval:gcs-flushing} we submit 50 million empty
tasks sequentially and monitor GCS memory consumption.  As expected, it grows
linearly with the number of tasks tracked and eventually reaches the memory
capacity of the system. At that point, the system becomes stalled and
the workload fails to finish within a reasonable amount of time.  With periodic
GCS flushing, we achieve two goals. First, the memory footprint is capped at a
user-configurable level (in the microbenchmark we employ an aggressive strategy
where consumed memory is kept as low as possible). Second, the flushing mechanism provides a natural way
to snapshot lineage to disk for long-running Ray applications.

\textbf{Recovering from task failures.}
In \figref{fig:fault_tolerance}, we demonstrate \System{}'s ability to
transparently recover from worker node failures and elastically scale, using the durable GCS lineage storage.
The workload, run on m4.xlarge instances, consists of linear chains of 100ms tasks submitted by the driver. As
nodes are removed (at 25s, 50s, 100s), the local schedulers reconstruct
previous results in the chain in order to continue execution. Overall \textit{per-node}
throughput remains stable throughout.

\textbf{Recovering from actor failures.}
By encoding actor method calls as stateful edges directly in the dependency graph, we
can reuse the same object reconstruction mechanism as in
\figref{fig:fault_tolerance} to provide transparent fault tolerance for \emph{stateful computation}.
\System{} additionally leverages user-defined checkpoint functions to
bound the reconstruction time for actors (Figure \ref{fig:actor_fault_tolerance}).
With minimal overhead, checkpointing enables only 500 methods to be re-executed, versus 10k re-executions without checkpointing.
In the future, we hope to further reduce actor reconstruction time, e.g., by allowing users to annotate methods that do not mutate state.

\begin{figure}[tb]
    \centering
    \begin{subfigure}[h]{0.228\textwidth}
        \centering
        \includegraphics[width=3.7cm]{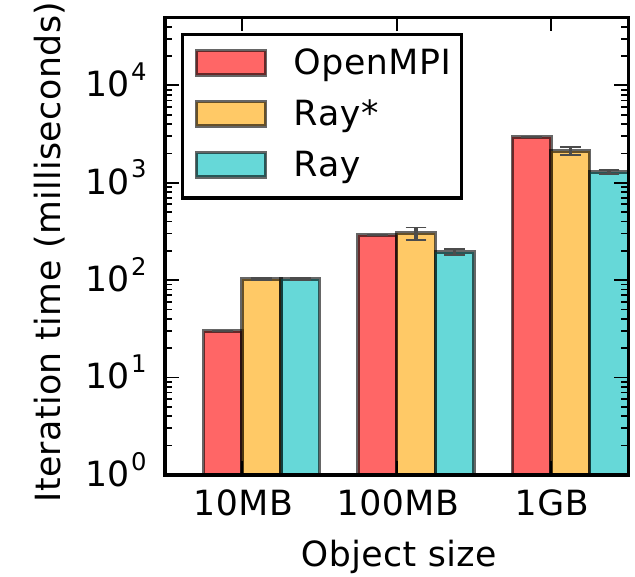}
        \caption{
            \System{} vs OpenMPI
        }
        \label{fig:allreduce_100}
    \end{subfigure}
    \hspace{.1cm}
    \begin{subfigure}[h]{0.228\textwidth}
        \centering
        \includegraphics[width=3.7cm]{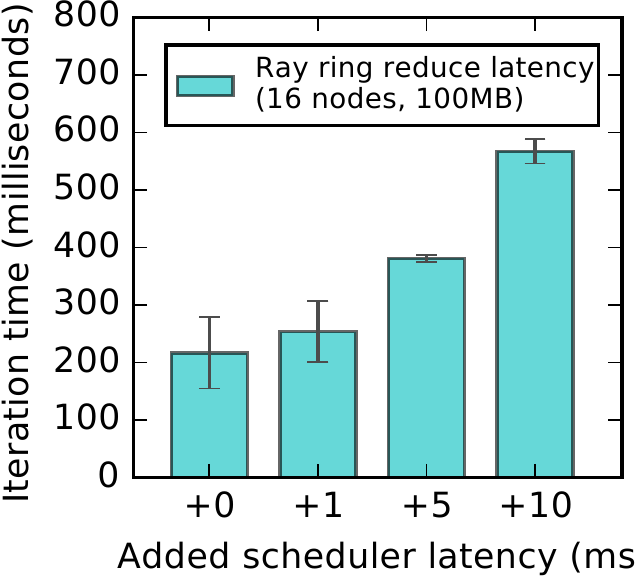}
        \caption{
            \System{} scheduler ablation
        }
        \label{fig:allreduce_ablation}
    \end{subfigure}
    \caption{
      \small{
      (a) Mean execution time of allreduce on 16 m4.16xl nodes. Each
        worker runs on a distinct node. \System{}* restricts \System{} to
        1 thread for sending and 1 thread for receiving.
      (b) \System{}'s low-latency scheduling is critical for allreduce.
    }
    }
\end{figure}

\noindent\hspace{1em}\textbf{Allreduce.}
Allreduce is a distributed communication primitive important to many machine learning workloads.
Here, we evaluate whether \System{} can natively support a ring allreduce \cite{allreducealgs} implementation with low enough overhead to match existing implementations~\cite{horovod}.
We find that \System{} completes allreduce across 16 nodes
on 100MB in $\sim$200ms and 1GB in $\sim$1200ms, surprisingly outperforming OpenMPI (v1.10), a
popular MPI implementation, by 1.5$\times$ and 2$\times$ respectively (\figref{fig:allreduce_100}).
We attribute \System{}'s performance to its use of multiple threads for network transfers, taking full advantage
of the 25Gbps connection between nodes on AWS, whereas OpenMPI sequentially sends and receives data on a single
thread~\cite{openmpi}.
For smaller objects, OpenMPI outperforms \System{} by switching to a lower overhead algorithm,
an optimization we plan to implement in the future.

\System{}'s scheduler performance is critical to implementing primitives such
as allreduce. In \figref{fig:allreduce_ablation}, we inject artificial task execution
delays and show that performance drops nearly 2$\times$ with just a few ms of extra latency.
Systems with centralized schedulers like Spark and CIEL typically
have scheduler overheads in the tens of milliseconds \cite{drizzle, ciel_thesis}, making such
workloads impractical. Scheduler \textit{throughput}
also becomes a bottleneck since the number of tasks required by ring reduce scales
quadratically with the number of participants.

\subsection{Building blocks}
\label{sec:building_blocks}

End-to-end applications (e.g., AlphaGo \cite{silver2016mastering}) require a
tight coupling of training, serving, and simulation. In this section, we isolate
each of these workloads to a setting that illustrates a typical RL application's requirements.
Due to a flexible programming model targeted to RL, and a system designed to support this programming
model, \System{} matches and sometimes exceeds the performance of dedicated systems for these individual workloads.

\begin{figure}
    \centering
    \includegraphics[width=3.1in,keepaspectratio]{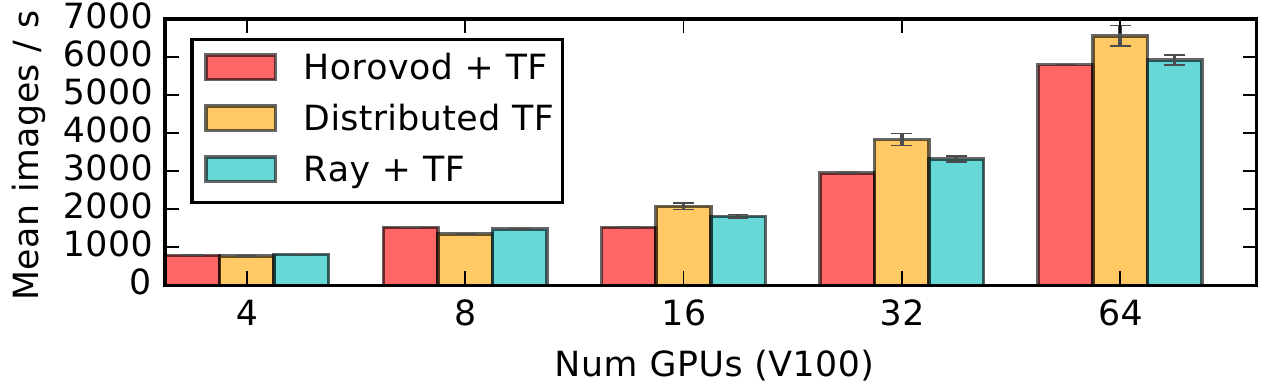}
    \caption{
    \small{
        Images per second reached when distributing the training of a
        ResNet-101 TensorFlow model (from the official TF benchmark).
        All experiments were run on p3.16xl instances connected by 25Gbps Ethernet, and
        workers allocated 4 GPUs per node as done in Horovod~\cite{horovod}.
        We note some measurement deviations from previously reported, likely
        due to hardware differences and
        recent TensorFlow performance improvements. We used
        OpenMPI 3.0, TF 1.8, and NCCL2 for all runs.
    }
    }
    \label{fig:sgd}
\end{figure}

\subsubsection{Distributed Training}
\label{sec:building_blocks:training}
We implement data-parallel synchronous SGD leveraging the \System{} actor
abstraction to represent model replicas. Model weights are synchronized
via allreduce~(\ref{sec:microbenchmarks}) or parameter server, both implemented on top of the \System{} API.

In Figure \ref{fig:sgd}, we evaluate the performance of the \System{} (synchronous)
parameter-server SGD implementation against state-of-the-art implementations
\cite{horovod}, using the same TensorFlow model and synthetic data generator for
each experiment. We compare only against TensorFlow-based systems to accurately
measure the overhead imposed by \System{}, rather than differences between the
deep learning frameworks themselves. In each iteration, model replica actors
compute gradients in parallel, send the gradients to a sharded parameter server,
then read the summed gradients from the parameter server for the next iteration.

Figure \ref{fig:sgd} shows that \System{} matches the performance of Horovod and
is within 10\% of distributed TensorFlow (in \texttt{distributed\_replicated} mode).
This is due to the ability to express the same application-level optimizations
found in these specialized systems in \System{}'s general-purpose API.
A key optimization is the pipelining of gradient computation, transfer, and
summation within a single iteration.
To overlap GPU computation with network transfer, we use a
custom TensorFlow operator to write tensors directly to \System{}'s
object store.

\subsubsection{Serving}
\label{sec:building_blocks:serving}
Model serving is an important component of end-to-end applications. \System{}
focuses primarily on the \textit{embedded} serving of models to simulators
running within the same dynamic task graph (e.g., within an RL application on
\System{}). In contrast, systems like Clipper
\cite{clipper} focus on serving predictions to external clients.

In this setting, low latency is critical for achieving high
utilization. To show this, in \tabref{tab:serving}
we compare the server throughput achieved using a \System{} actor
to serve a policy versus using the open source Clipper system over REST. Here, both client and server processes are co-located on the same machine (a p3.8xlarge instance).
This is often the case for RL applications but not for the general web serving workloads addressed by systems like Clipper.
Due to its low-overhead serialization and shared memory abstractions,
\System{} achieves an order of magnitude higher throughput for a small fully connected policy model that takes in a large input and is also faster on a more expensive residual network policy model, similar to one used in AlphaGo Zero, that takes smaller input.

\begin{table}[tb]
\begin{center}
\begin{small}
\begin{tabular}{| c | c | c |}
    \hline
    {\bf System}& {\bf Small Input} & {\bf Larger Input}\\\hline
    Clipper     & 4400 $\pm$ 15 states/sec & 290 $\pm$ 1.3 states/sec \\\hline
    \System{}   & 6200 $\pm$ 21 states/sec  & 6900 $\pm$ 150 states/sec \\\hline
\end{tabular}
\end{small}
\end{center}
\caption{\small{Throughput comparisons for Clipper~\cite{clipper}, a dedicated serving system, and
\System{} for two embedded serving workloads. We use a residual network and a small fully connected network, taking 10ms and 5ms to evaluate,
respectively. The server is queried by clients that each send states of size 4KB and 100KB respectively in batches of 64.}}
\label{tab:serving}

\end{table}

\subsubsection{Simulation}
\label{sec:building_blocks:simulation}
Simulators used in RL produce results with variable
lengths (``timesteps'') that, due to the tight loop with training, must be used as soon as they
are available.  The task heterogeneity and timeliness requirements make
simulations hard to support efficiently in BSP-style systems.  To demonstrate,
we compare (1) an MPI implementation that submits $3n$ parallel simulation runs
on $n$ cores in 3 rounds, with a global barrier between rounds\footnote{Note
  that experts \emph{can} use MPI's asynchronous primitives to get around
  barriers---at the expense of increased program complexity ---we nonetheless
  chose such an implementation to simulate BSP.}, to (2) a \System{} program
that issues the same $3n$ tasks while concurrently gathering simulation
results back to the driver. Table \ref{table:simulation} shows that both systems
scale well, yet \System{} achieves up to 1.8$\times$ throughput.
This motivates a programming model that can dynamically spawn and collect the results of fine-grained simulation tasks.

\begin{table}[tbh]
\begin{center}
\begin{footnotesize}
\begin{tabular}{| c | c | c | c |}
\hline
{\bf System, programming model} & 1 CPU & 16 CPUs & 256 CPUs \\\hline
MPI, bulk synchronous           & 22.6K & 208K & 2.16M \\\hline
\System{}, asynchronous tasks   & 22.3K & 290K & 4.03M \\ \hline
\end{tabular}
\end{footnotesize}
\end{center}
\caption{\small{Timesteps per second for the Pendulum-v0 simulator in OpenAI
    Gym~\cite{brockman2016openai}.  \System{} allows for
    better utilization when running heterogeneous simulations at scale.}}
\label{table:simulation}
\end{table}

\subsection{RL Applications}
\label{sec:exp_rl}

Without a system that can tightly couple the training, simulation, and serving steps, reinforcement learning algorithms today are implemented as one-off solutions that make it difficult to incorporate optimizations that, for example, require a different computation structure or that utilize different architectures.
Consequently, with implementations of two representative reinforcement learning applications in \System{}, we are able to match and even outperform custom systems built specifically for these algorithms.
The primary reason is the flexibility of \System{}'s programming model, which can express application-level optimizations that would require substantial engineering effort to port to custom-built systems, but are transparently supported by \System{}'s dynamic task graph execution engine.

\subsubsection{Evolution Strategies}
\label{sec:exp_rl:es}
To evaluate \System{} on large-scale RL workloads,
we implement the evolution strategies (ES) algorithm
and compare to the reference implementation \cite{salimans2017evolution}---a
system specially built for this algorithm that relies on Redis for messaging and
low-level multiprocessing libraries for data-sharing. The algorithm periodically
broadcasts a new policy to a pool of workers and aggregates the results of
roughly 10000 tasks (each performing 10 to 1000 simulation steps).

\begin{figure}[bt]
    \centering
    \begin{subfigure}[b]{0.228\textwidth}
        \centering
        \includegraphics[width=3.5cm]{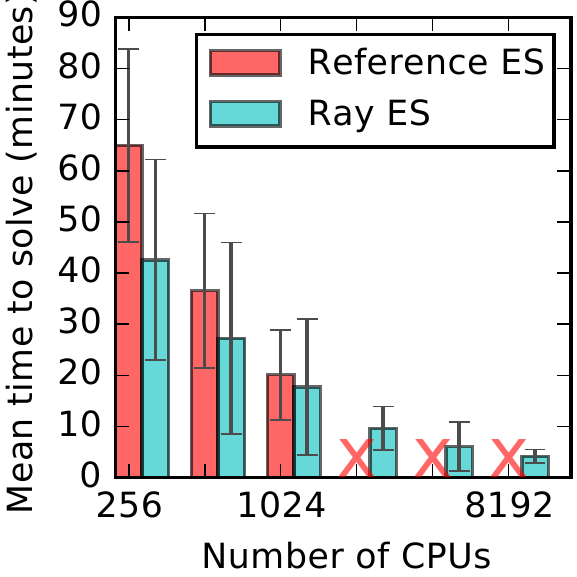}
        \caption{
            Evolution Strategies
        }
        \label{fig:es_scaling}
    \end{subfigure}
    \hspace{.1cm}
    \begin{subfigure}[b]{0.228\textwidth}
        \centering
        \includegraphics[width=3.8cm]{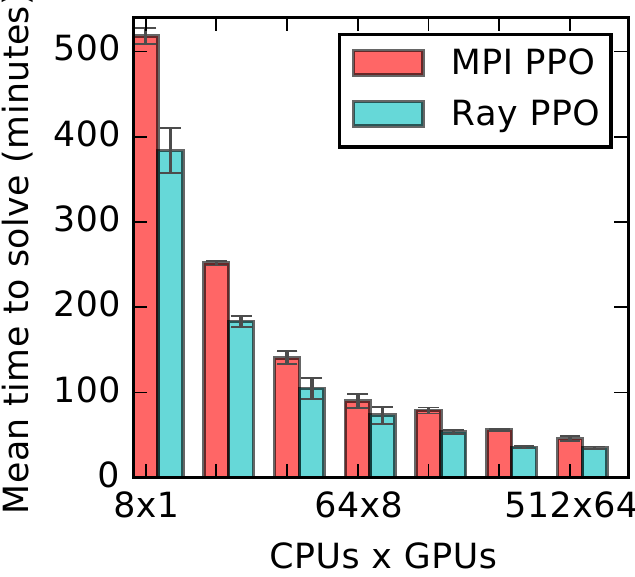}
        \caption{
            PPO
        }
        \label{fig:humanoid_ppo}
    \end{subfigure}
    \caption{
      \small{
    	Time to reach a score of 6000 in the Humanoid-v1 task \citep{brockman2016openai}.
    	\textbf{(a)} The \System{} ES implementation scales well to
        8192 cores and achieves a median time of 3.7 minutes,
        over twice as fast as the best published result.
        The special-purpose system failed to run beyond 1024 cores.
        ES is
        faster than PPO on this benchmark, but shows greater runtime variance.
    	\textbf{(b)} The \System{} PPO implementation outperforms a specialized MPI
        implementation \cite{openaibaselines} with fewer GPUs, at a fraction of the
        cost. The MPI implementation required 1 GPU for every 8 CPUs, whereas the
        \System{} version required at most 8 GPUs (and never more than 1 GPU per 8
        CPUs).
    }
    }
	\vspace{-0.2in}
\end{figure}

As shown in \figref{fig:es_scaling}, an implementation on \System{} scales to 8192 cores.
Doubling the cores available yields an average completion time speedup of 1.6$\times$.
Conversely, the special-purpose system fails to complete at 2048 cores, where the work in the system exceeds the processing capacity of the application driver.
To avoid this issue, the \System{} implementation uses an aggregation tree of actors, reaching a median time of 3.7 minutes, more than twice as fast as the best published result (10 minutes).

Initial parallelization of a serial implementation using \System{} required modifying
only 7 lines of code. Performance improvement through hierarchical
aggregation was easy to realize with \System{}'s support for nested tasks and
actors.
In contrast, the reference implementation had several hundred lines of code
dedicated to a protocol for communicating tasks and data between workers, and
would require further engineering to support optimizations like hierarchical
aggregation.

\subsubsection{Proximal Policy Optimization}
\label{sec:exp_rl:ppo}
We
implement Proximal Policy Optimization (PPO) \cite{schulman2017proximal} in
\System{} and compare to a highly-optimized reference implementation
\cite{openaibaselines} that uses OpenMPI communication primitives.
The algorithm is an asynchronous scatter-gather, where new tasks are assigned
to simulation actors as they return rollouts to the driver. Tasks are submitted until
320000 simulation steps are collected (each task produces between 10 and
1000 steps). The policy update performs 20 steps of SGD with a batch size of
32768. The model parameters in this example are roughly 350KB.
These experiments were run using p2.16xlarge (GPU) and m4.16xlarge (high CPU) instances.

 As shown in \figref{fig:humanoid_ppo}, the \System{} implementation outperforms
the optimized MPI implementation in all experiments, while using a
fraction of the GPUs. The reason is that \System{} is
heterogeneity-aware and allows the user to utilize asymmetric architectures by
expressing resource requirements at the granularity of a task or actor.
The \System{} implementation can then leverage TensorFlow's single-process
multi-GPU support and can pin objects in GPU memory when possible. This
optimization cannot be easily ported to MPI due to the need to asynchronously
gather rollouts to a single GPU process. Indeed, \cite{openaibaselines} includes
two custom implementations of PPO, one using MPI for large clusters and one that
is optimized for GPUs but that is restricted to a single node. \System{} allows
for an implementation suitable for both scenarios.

\System{}'s ability to handle resource heterogeneity also decreased PPO's cost by a
factor of 4.5 \cite{ec2pricing}, since CPU-only tasks can be scheduled on
cheaper high-CPU instances. In contrast, MPI applications often exhibit
symmetric architectures, in which all processes run the same code and require
identical resources, in this case preventing the use of CPU-only machines for
scale-out. Furthermore, the MPI implementation requires on-demand instances
since it does not transparently handle failure. Assuming 4$\times$ cheaper spot
instances, {\em \System{}'s fault tolerance and resource-aware scheduling
together cut costs by 18$\times$}.

\section{Related Work}
\label{sec:prevwork}

\looseness=-1
\hspace{1em}\textbf{Dynamic task graphs.}
{\System} is closely related to CIEL~\cite{ciel} and Dask~\citep{dask-scipy15}.
All three support dynamic task graphs with nested tasks and implement the futures abstraction.
CIEL also provides lineage-based fault tolerance, while Dask, like {\System}, fully integrates with Python.
However, {\System} differs in two aspects that have important performance consequences.
First, {\System} extends the task model with an actor abstraction.
This is necessary for efficient stateful computation in distributed training and serving, to keep the model data collocated with the computation.
Second, {\System} employs a fully distributed and decoupled control plane and scheduler, instead of relying on a single master storing all metadata.
This is critical for efficiently supporting primitives like allreduce without system modification.
At peak performance for 100MB on 16 nodes, allreduce on {\System} (Section \ref{sec:microbenchmarks}) submits 32 rounds of 16 tasks in 200ms.
Meanwhile, Dask reports a maximum scheduler throughput of 3k tasks/s on 512 cores~\citep{dask-benchmarks}.
With a centralized scheduler, each round of allreduce would then incur a minimum of $\sim$5ms of scheduling delay, translating to up to $2\times$ worse completion time (Figure \ref{fig:allreduce_ablation}).
Even with a decentralized scheduler, coupling the control plane information with the scheduler leaves the latter on the critical path for data transfer, adding an extra roundtrip to every round of allreduce.

\textbf{Dataflow systems.}
Popular dataflow systems, such as MapReduce~\citep{mapreduce}, Spark~\citep{spark-cacm16}, and Dryad~\citep{dryad} have widespread adoption for analytics and ML workloads, but their computation model is too restrictive for a fine-grained and dynamic simulation workload.
Spark and MapReduce implement the BSP execution model, which assumes that tasks within the same stage perform the same computation and take roughly the same amount of time.
Dryad relaxes this restriction but lacks support for dynamic task graphs.
Furthermore, none of these systems provide an actor abstraction, nor implement a distributed scalable control plane and scheduler.
Finally, Naiad~\citep{naiad} is a dataflow system that provides improved scalability for some workloads, but only supports static task graphs.

\textbf{Machine learning frameworks.}
TensorFlow~\citep{tensorflow-osdi16} and MXNet~\citep{mxnet-learningsys} target deep learning workloads and efficiently leverage both CPUs and GPUs.
While they achieve great performance for training workloads consisting of static DAGs of linear algebra operations, they have limited support for the more general computation required to tightly couple training with simulation and embedded serving.
TensorFlow Fold~\citep{looks2017deep} provides some support for dynamic task graphs, as well as MXNet through its internal C++ APIs, but neither fully supports the ability to modify the DAG during execution in response to task progress, task completion times, or faults.
TensorFlow and MXNet in principle achieve generality by allowing the programmer to simulate low-level message-passing and synchronization primitives, but the pitfalls and user experience in this case are similar to those of MPI.
OpenMPI~\citep{openmpi} can achieve high performance, but it is relatively hard to program as it requires explicit coordination to handle heterogeneous and dynamic task graphs.
Furthermore, it forces the programmer to explicitly handle fault tolerance.

\textbf{Actor systems.} Orleans~\citep{bykov2011orleans} and Akka~\citep{akka}
are two actor frameworks well suited to developing highly available and
concurrent distributed systems. However, compared to \System{}, they provide less support
for recovery from data loss.  To recover \emph{stateful actors}, the
Orleans developer must explicitly checkpoint actor state and intermediate
responses.  \emph{Stateless actors} in Orleans can be replicated for scale-out, and could therefore act as tasks, but unlike in
{\System}, they have no lineage. Similarly, while Akka explicitly supports persisting actor state across failures, it does not provide efficient fault tolerance for
\emph{stateless computation} (i.e.,
tasks). For message delivery, Orleans provides at-least-once and Akka
provides at-most-once semantics. In contrast, {\System} provides transparent
fault tolerance and exactly-once semantics, as each method call is logged in the
GCS and both arguments and results are immutable.  We find that in practice
these limitations do not affect the performance of our applications.
Erlang~\citep{armstrong1993concurrent} and C++ Actor Framework~\citep{charousset2013native}, two other actor-based systems, have similarly limited support for fault tolerance.

\textbf{Global control store and scheduling.}
The concept of logically centralizing the control plane has been previously
proposed in software defined networks (SDNs) \citep{ethane-sdn}, distributed
file systems (e.g., GFS~\citep{ghemawat2003google}), resource management
(e.g., Omega~\citep{omega-eurosys13}), and distributed frameworks
(e.g., MapReduce~\citep{mapreduce}, BOOM~\citep{alvaro2010boom}), to name a few.
{\System} draws inspiration from these pioneering efforts, but provides
significant improvements. In contrast with SDNs, BOOM, and GFS, {\System} decouples the storage of the
control plane information (e.g., GCS) from the logic implementation
(e.g., schedulers). This allows both storage and computation layers to scale
independently, which is key to achieving our scalability targets. Omega uses a
distributed architecture in which schedulers coordinate via globally shared
state. To this architecture, {\System} adds global schedulers to balance load
across local schedulers, and targets ms-level, not second-level, task
scheduling.

{\System} implements a unique distributed bottom-up scheduler that is
horizontally scalable, and can handle dynamically constructed task graphs.
Unlike {\System}, most existing cluster computing
systems~\citep{mapreduce,spark-nsdi12,ciel} use a centralized scheduler
architecture. While Sparrow \citep{sparrow} is decentralized, its schedulers
make independent decisions, limiting the possible scheduling policies, and all
tasks of a job are handled by the same global scheduler. Mesos~\citep{mesos}
implements a two-level hierarchical scheduler, but its top-level scheduler manages
frameworks, not individual tasks. Canary~\citep{qu2016canary} achieves impressive performance by
having each scheduler instance handle a portion of the task graph, but does not
handle dynamic computation graphs.

Cilk~\citep{work-stealing} is a parallel programming language whose work-stealing scheduler achieves provably efficient load-balancing for dynamic task graphs.
However, with no central coordinator like {\System}'s global scheduler, this fully parallel design is also difficult to extend to support data locality and resource heterogeneity in a distributed setting.

\section{Discussion and Experiences}
\label{sec:conclusion}

Building {\System} has been a long journey. It started two years ago with a
Spark library to perform distributed training and simulations.
However, the relative inflexibility of the BSP model, the  high per-task
overhead, and the lack of an actor abstraction led us to develop a new system.
Since we released {\System} roughly one year ago, several hundreds of people
have used it and several companies are running it in production.
Here we discuss our experience developing and using {\System}, and
some early user feedback.

{\bf API.} In designing the API, we have emphasized minimalism.
Initially we started with a basic \emph{task} abstraction. Later, we added the
\textbf{wait()} primitive to accommodate rollouts with heterogeneous durations
and the \emph{actor} abstraction to accommodate third-party simulators and
amortize the overhead of expensive initializations. While the resulting API is
relatively low-level, it has proven both powerful and simple to use. We have
already used this API to implement many state-of-the-art RL algorithms on top of
{\System}, including
A3C~\cite{mnih2016asynchronous},
PPO~\cite{schulman2017proximal},
DQN~\cite{mnih2015human},
ES~\cite{salimans2017evolution},
DDPG~\cite{silver2014deterministic},
and Ape-X~\cite{horgan2018distributed}.
In most cases it took us just a few
tens of lines of code to port these algorithms to {\System}.
Based on early user feedback, we
are considering enhancing the API
to include higher level primitives and libraries, which
could also inform scheduling decisions.

\textbf{Limitations.} Given the workload generality, specialized optimizations
are hard. For example, we must make scheduling decisions without full knowledge
of the computation graph. Scheduling optimizations in {\System} might require
more complex runtime profiling. In addition, storing lineage for each task
requires the implementation of garbage collection policies to bound storage
costs in the GCS, a feature we are actively developing.

\textbf{Fault tolerance.} We are often asked if fault tolerance is really needed
for AI applications. After all, due to the statistical nature of many AI
algorithms, one could simply ignore failed rollouts. Based on our experience,
our answer is ``yes''. First, the ability to ignore failures makes applications
much easier to write and reason about. Second, our particular implementation of
fault tolerance via deterministic replay dramatically simplifies debugging as it
allows us to easily reproduce most errors. This is particularly important since,
due to their stochasticity, AI algorithms are notoriously hard to debug. Third,
fault tolerance helps save money since it allows us to run on cheap resources
like spot instances on AWS.
Of course, this comes at the price of some overhead. However, we found this
overhead to be minimal for our target workloads.

\textbf{GCS and Horizontal Scalability.} The GCS dramatically simplified \System{}
development and debugging. It enabled us to query
the entire system state while debugging \System{} itself, instead of having to manually expose
internal component state. In addition, the GCS is also the backend for our timeline visualization tool, used for application-level debugging.

The GCS was also instrumental to \System{}'s horizontal scalability. In \secref{sec:experiments}, we were able to scale
by adding more shards whenever the GCS became a bottleneck. The GCS also
enabled the global scheduler to scale by simply adding more replicas.
Due to these advantages, we believe that centralizing control state will be a
key design component of future distributed systems.

\section{Conclusion}
\label{sec:conclusion}

 No general-purpose system today can efficiently support the tight loop of training, serving, and simulation.
 To express these core building blocks and meet the demands of emerging AI applications,
 \System{} unifies task-parallel and actor programming models in a single dynamic task graph
 and employs a scalable architecture enabled by the global control store and a bottom-up distributed scheduler.
 The programming flexibility, high throughput, and low latencies simultaneously achieved by this architecture
 is particularly important for emerging artificial intelligence workloads,
 which produce tasks diverse in their resource requirements, duration, and functionality.
 Our evaluation demonstrates linear scalability up to 1.8 million tasks per
 second, transparent fault tolerance, and substantial performance improvements on
 several contemporary RL workloads. Thus, {\System} provides a powerful
 combination of flexibility, performance, and ease of use for the development of
 future AI applications.

\section{Acknowledgments}
\label{sec:acks}
This research is supported in part by
NSF CISE Expeditions Award CCF-1730628 and gifts from
Alibaba, Amazon Web Services, Ant Financial, Arm, CapitalOne, Ericsson, Facebook, Google, Huawei, Intel,
Microsoft, Scotiabank, Splunk and VMware as well as by NSF grant DGE-1106400.
We are grateful to our anonymous reviewers and our shepherd, Miguel Castro, for thoughtful feedback,
which helped improve the quality of this paper.

{\footnotesize \bibliographystyle{acm}
\bibliography{ray}}

\begin{thebibliography}{10}

\bibitem{akka}
Akka.
\newblock \url{https://akka.io/}.

\bibitem{arrow}
{Apache Arrow}.
\newblock \url{https://arrow.apache.org/}.

\bibitem{dask-benchmarks}
{Dask Benchmarks}.
\newblock \url{http://matthewrocklin.com/blog/work/2017/07/03/scaling}.

\bibitem{ec2pricing}
{EC2 Instance Pricing}.
\newblock \url{https://aws.amazon.com/ec2/pricing/on-demand/}.

\bibitem{openaibaselines}
{OpenAI Baselines}: high-quality implementations of reinforcement learning
  algorithms.
\newblock \url{https://github.com/openai/baselines}.

\bibitem{tensorflow-serving}
{TensorFlow Serving}.
\newblock \url{https://www.tensorflow.org/serving/}.

\bibitem{tensorflow-osdi16}
{\sc Abadi, M., Barham, P., Chen, J., Chen, Z., Davis, A., Dean, J., Devin, M.,
  Ghemawat, S., Irving, G., Isard, M., et~al.}
\newblock {TensorFlow}: A system for large-scale machine learning.
\newblock In {\em Proceedings of the 12th USENIX Symposium on Operating Systems
  Design and Implementation (OSDI). Savannah, Georgia, USA\/} (2016).

\bibitem{agarwal2016multiworld}
{\sc Agarwal, A., Bird, S., Cozowicz, M., Hoang, L., Langford, J., Lee, S., Li,
  J., Melamed, D., Oshri, G., Ribas, O., Sen, S., and Slivkins, A.}
\newblock A multiworld testing decision service.
\newblock {\em arXiv preprint arXiv:1606.03966\/} (2016).

\bibitem{alvaro2010boom}
{\sc Alvaro, P., Condie, T., Conway, N., Elmeleegy, K., Hellerstein, J.~M., and
  Sears, R.}
\newblock {BOOM} {A}nalytics: exploring data-centric, declarative programming
  for the cloud.
\newblock In {\em Proceedings of the 5th European conference on Computer
  systems\/} (2010), ACM, pp.~223--236.

\bibitem{armstrong1993concurrent}
{\sc Armstrong, J., Virding, R., Wikstr{\"o}m, C., and Williams, M.}
\newblock Concurrent programming in {ERLANG}.

\bibitem{beattie2016deepmind}
{\sc Beattie, C., Leibo, J.~Z., Teplyashin, D., Ward, T., Wainwright, M.,
  K{\"u}ttler, H., Lefrancq, A., Green, S., Vald{\'e}s, V., Sadik, A., et~al.}
\newblock {DeepMind Lab}.
\newblock {\em arXiv preprint arXiv:1612.03801\/} (2016).

\bibitem{work-stealing}
{\sc Blumofe, R.~D., and Leiserson, C.~E.}
\newblock Scheduling multithreaded computations by work stealing.
\newblock {\em J. ACM 46}, 5 (Sept. 1999), 720--748.

\bibitem{brockman2016openai}
{\sc Brockman, G., Cheung, V., Pettersson, L., Schneider, J., Schulman, J.,
  Tang, J., and Zaremba, W.}
\newblock {OpenAI} gym.
\newblock {\em arXiv preprint arXiv:1606.01540\/} (2016).

\bibitem{bykov2011orleans}
{\sc Bykov, S., Geller, A., Kliot, G., Larus, J.~R., Pandya, R., and Thelin,
  J.}
\newblock Orleans: {C}loud computing for everyone.
\newblock In {\em Proceedings of the 2nd ACM Symposium on Cloud Computing\/}
  (2011), ACM, p.~16.

\bibitem{flink}
{\sc Carbone, P., Ewen, S., F\'{o}ra, G., Haridi, S., Richter, S., and Tzoumas,
  K.}
\newblock State management in {Apache Flink}: Consistent stateful distributed
  stream processing.
\newblock {\em Proc. VLDB Endow. 10}, 12 (Aug. 2017), 1718--1729.

\bibitem{ethane-sdn}
{\sc Casado, M., Freedman, M.~J., Pettit, J., Luo, J., McKeown, N., and
  Shenker, S.}
\newblock Ethane: Taking control of the enterprise.
\newblock {\em SIGCOMM Comput. Commun. Rev. 37}, 4 (Aug. 2007), 1--12.

\bibitem{charousset2013native}
{\sc Charousset, D., Schmidt, T.~C., Hiesgen, R., and W{\"a}hlisch, M.}
\newblock Native actors: A scalable software platform for distributed,
  heterogeneous environments.
\newblock In {\em Proceedings of the 2013 workshop on Programming based on
  actors, agents, and decentralized control\/} (2013), ACM, pp.~87--96.

\bibitem{mxnet-learningsys}
{\sc Chen, T., Li, M., Li, Y., Lin, M., Wang, N., Wang, M., Xiao, T., Xu, B.,
  Zhang, C., and Zhang, Z.}
\newblock {MXNet}: A flexible and efficient machine learning library for
  heterogeneous distributed systems.
\newblock In {\em NIPS Workshop on Machine Learning Systems (LearningSys'16)\/}
  (2016).

\bibitem{clipper}
{\sc Crankshaw, D., Wang, X., Zhou, G., Franklin, M.~J., Gonzalez, J.~E., and
  Stoica, I.}
\newblock Clipper: A low-latency online prediction serving system.
\newblock In {\em 14th {USENIX} Symposium on Networked Systems Design and
  Implementation ({NSDI} 17)\/} (Boston, MA, 2017), {USENIX} Association,
  pp.~613--627.

\bibitem{mapreduce}
{\sc Dean, J., and Ghemawat, S.}
\newblock {MapReduce}: Simplified data processing on large clusters.
\newblock {\em Commun. ACM 51}, 1 (Jan. 2008), 107--113.

\bibitem{dataflow74}
{\sc Dennis, J.~B., and Misunas, D.~P.}
\newblock A preliminary architecture for a basic data-flow processor.
\newblock In {\em Proceedings of the 2Nd Annual Symposium on Computer
  Architecture\/} (New York, NY, USA, 1975), ISCA '75, ACM, pp.~126--132.

\bibitem{openmpi}
{\sc Gabriel, E., Fagg, G.~E., Bosilca, G., Angskun, T., Dongarra, J.~J.,
  Squyres, J.~M., Sahay, V., Kambadur, P., Barrett, B., Lumsdaine, A., Castain,
  R.~H., Daniel, D.~J., Graham, R.~L., and Woodall, T.~S.}
\newblock Open {MPI}: Goals, concept, and design of a next generation {MPI}
  implementation.
\newblock In {\em Proceedings, 11th European PVM/MPI Users' Group Meeting\/}
  (Budapest, Hungary, September 2004), pp.~97--104.

\bibitem{ghemawat2003google}
{\sc Ghemawat, S., Gobioff, H., and Leung, S.-T.}
\newblock The {G}oogle file system.
\newblock 29--43.

\bibitem{graphx}
{\sc Gonzalez, J.~E., Xin, R.~S., Dave, A., Crankshaw, D., Franklin, M.~J., and
  Stoica, I.}
\newblock {GraphX}: Graph processing in a distributed dataflow framework.
\newblock In {\em Proceedings of the 11th USENIX Conference on Operating
  Systems Design and Implementation\/} (Berkeley, CA, USA, 2014), OSDI'14,
  USENIX Association, pp.~599--613.

\bibitem{GuHolLilLev17}
{\sc Gu*, S., Holly*, E., Lillicrap, T., and Levine, S.}
\newblock Deep reinforcement learning for robotic manipulation with
  asynchronous off-policy updates.
\newblock In {\em IEEE International Conference on Robotics and Automation
  (ICRA 2017)\/} (2017).

\bibitem{mesos}
{\sc Hindman, B., Konwinski, A., Zaharia, M., Ghodsi, A., Joseph, A.~D., Katz,
  R., Shenker, S., and Stoica, I.}
\newblock Mesos: A platform for fine-grained resource sharing in the data
  center.
\newblock In {\em Proceedings of the 8th USENIX Conference on Networked Systems
  Design and Implementation\/} (Berkeley, CA, USA, 2011), NSDI'11, USENIX
  Association, pp.~295--308.

\bibitem{horgan2018distributed}
{\sc Horgan, D., Quan, J., Budden, D., Barth-Maron, G., Hessel, M., van
  Hasselt, H., and Silver, D.}
\newblock Distributed prioritized experience replay.
\newblock {\em International Conference on Learning Representations\/} (2018).

\bibitem{dryad}
{\sc Isard, M., Budiu, M., Yu, Y., Birrell, A., and Fetterly, D.}
\newblock Dryad: Distributed data-parallel programs from sequential building
  blocks.
\newblock In {\em Proceedings of the 2nd ACM SIGOPS/EuroSys European Conference
  on Computer Systems 2007\/} (New York, NY, USA, 2007), EuroSys '07, ACM,
  pp.~59--72.

\bibitem{jia2014caffe}
{\sc Jia, Y., Shelhamer, E., Donahue, J., Karayev, S., Long, J., Girshick, R.,
  Guadarrama, S., and Darrell, T.}
\newblock Caffe: Convolutional architecture for fast feature embedding.
\newblock {\em arXiv preprint arXiv:1408.5093\/} (2014).

\bibitem{jordan2015machine}
{\sc Jordan, M.~I., and Mitchell, T.~M.}
\newblock Machine learning: Trends, perspectives, and prospects.
\newblock {\em Science 349}, 6245 (2015), 255--260.

\bibitem{storm}
{\sc Leibiusky, J., Eisbruch, G., and Simonassi, D.}
\newblock {\em Getting Started with Storm}.
\newblock O'Reilly Media, Inc., 2012.

\bibitem{param-server}
{\sc Li, M., Andersen, D.~G., Park, J.~W., Smola, A.~J., Ahmed, A., Josifovski,
  V., Long, J., Shekita, E.~J., and Su, B.-Y.}
\newblock Scaling distributed machine learning with the parameter server.
\newblock In {\em Proceedings of the 11th USENIX Conference on Operating
  Systems Design and Implementation\/} (Berkeley, CA, USA, 2014), OSDI'14,
  pp.~583--598.

\bibitem{looks2017deep}
{\sc Looks, M., Herreshoff, M., Hutchins, D., and Norvig, P.}
\newblock Deep learning with dynamic computation graphs.
\newblock {\em arXiv preprint arXiv:1702.02181\/} (2017).

\bibitem{graphlab}
{\sc Low, Y., Gonzalez, J., Kyrola, A., Bickson, D., Guestrin, C., and
  Hellerstein, J.}
\newblock {GraphLab}: A new framework for parallel machine learning.
\newblock In {\em Proceedings of the Twenty-Sixth Conference on Uncertainty in
  Artificial Intelligence\/} (Arlington, Virginia, United States, 2010),
  UAI'10, pp.~340--349.

\bibitem{pregel}
{\sc Malewicz, G., Austern, M.~H., Bik, A.~J., Dehnert, J.~C., Horn, I.,
  Leiser, N., and Czajkowski, G.}
\newblock Pregel: A system for large-scale graph processing.
\newblock In {\em Proceedings of the 2010 ACM SIGMOD International Conference
  on Management of Data\/} (New York, NY, USA, 2010), SIGMOD '10, ACM,
  pp.~135--146.

\bibitem{mnih2016asynchronous}
{\sc Mnih, V., Badia, A.~P., Mirza, M., Graves, A., Lillicrap, T.~P., Harley,
  T., Silver, D., and Kavukcuoglu, K.}
\newblock Asynchronous methods for deep reinforcement learning.
\newblock In {\em International Conference on Machine Learning\/} (2016).

\bibitem{mnih2015human}
{\sc Mnih, V., Kavukcuoglu, K., Silver, D., Rusu, A.~A., Veness, J., Bellemare,
  M.~G., Graves, A., Riedmiller, M., Fidjeland, A.~K., Ostrovski, G., et~al.}
\newblock Human-level control through deep reinforcement learning.
\newblock {\em Nature 518}, 7540 (2015), 529--533.

\bibitem{ciel_thesis}
{\sc Murray, D.}
\newblock {\em A Distributed Execution Engine Supporting Data-dependent Control
  Flow}.
\newblock University of Cambridge, 2012.

\bibitem{naiad}
{\sc Murray, D.~G., McSherry, F., Isaacs, R., Isard, M., Barham, P., and Abadi,
  M.}
\newblock Naiad: A timely dataflow system.
\newblock In {\em Proceedings of the Twenty-Fourth ACM Symposium on Operating
  Systems Principles\/} (New York, NY, USA, 2013), SOSP '13, ACM, pp.~439--455.

\bibitem{ciel}
{\sc Murray, D.~G., Schwarzkopf, M., Smowton, C., Smith, S., Madhavapeddy, A.,
  and Hand, S.}
\newblock {CIEL}: A universal execution engine for distributed data-flow
  computing.
\newblock In {\em Proceedings of the 8th USENIX Conference on Networked Systems
  Design and Implementation\/} (Berkeley, CA, USA, 2011), NSDI'11, USENIX
  Association, pp.~113--126.

\bibitem{gorila}
{\sc Nair, A., Srinivasan, P., Blackwell, S., Alcicek, C., Fearon, R., Maria,
  A.~D., Panneershelvam, V., Suleyman, M., Beattie, C., Petersen, S., Legg, S.,
  Mnih, V., Kavukcuoglu, K., and Silver, D.}
\newblock Massively parallel methods for deep reinforcement learning, 2015.

\bibitem{ng2006autonomous}
{\sc Ng, A., Coates, A., Diel, M., Ganapathi, V., Schulte, J., Tse, B., Berger,
  E., and Liang, E.}
\newblock Autonomous inverted helicopter flight via reinforcement learning.
\newblock {\em Experimental Robotics IX\/} (2006), 363--372.

\bibitem{nishihara2017real}
{\sc Nishihara, R., Moritz, P., Wang, S., Tumanov, A., Paul, W.,
  Schleier-Smith, J., Liaw, R., Niknami, M., Jordan, M.~I., and Stoica, I.}
\newblock Real-time machine learning: The missing pieces.
\newblock In {\em Workshop on Hot Topics in Operating Systems\/} (2017).

\bibitem{dota}
{\sc OpenAI}.
\newblock {OpenAI Dota 2 1v1 bot}.
\newblock \url{https://openai.com/the-international/}, 2017.

\bibitem{sparrow}
{\sc Ousterhout, K., Wendell, P., Zaharia, M., and Stoica, I.}
\newblock Sparrow: Distributed, low latency scheduling.
\newblock In {\em Proceedings of the Twenty-Fourth ACM Symposium on Operating
  Systems Principles\/} (New York, NY, USA, 2013), SOSP '13, ACM, pp.~69--84.

\bibitem{pytorch}
{\sc Paszke, A., Gross, S., Chintala, S., Chanan, G., Yang, E., DeVito, Z.,
  Lin, Z., Desmaison, A., Antiga, L., and Lerer, A.}
\newblock Automatic differentiation in {PyTorch}.

\bibitem{qu2016canary}
{\sc Qu, H., Mashayekhi, O., Terei, D., and Levis, P.}
\newblock Canary: A scheduling architecture for high performance cloud
  computing.
\newblock {\em arXiv preprint arXiv:1602.01412\/} (2016).

\bibitem{dask-scipy15}
{\sc Rocklin, M.}
\newblock Dask: Parallel computation with blocked algorithms and task
  scheduling.
\newblock In {\em Proceedings of the 14th Python in Science Conference\/}
  (2015), K.~Huff and J.~Bergstra, Eds., pp.~130 -- 136.

\bibitem{salimans2017evolution}
{\sc Salimans, T., Ho, J., Chen, X., and Sutskever, I.}
\newblock Evolution strategies as a scalable alternative to reinforcement
  learning.
\newblock {\em arXiv preprint arXiv:1703.03864\/} (2017).

\bibitem{redis2009}
{\sc Sanfilippo, S.}
\newblock {Redis: An open source, in-memory data structure store}.
\newblock \url{https://redis.io/}, 2009.

\bibitem{schulman2017proximal}
{\sc Schulman, J., Wolski, F., Dhariwal, P., Radford, A., and Klimov, O.}
\newblock Proximal policy optimization algorithms.
\newblock {\em arXiv preprint arXiv:1707.06347\/} (2017).

\bibitem{omega-eurosys13}
{\sc Schwarzkopf, M., Konwinski, A., Abd-El-Malek, M., and Wilkes, J.}
\newblock Omega: Flexible, scalable schedulers for large compute clusters.
\newblock In {\em Proceedings of the 8th ACM European Conference on Computer
  Systems\/} (New York, NY, USA, 2013), EuroSys '13, ACM, pp.~351--364.

\bibitem{horovod}
{\sc Sergeev, A., and Del~Balso, M.}
\newblock Horovod: fast and easy distributed deep learning in tensorflow.
\newblock {\em arXiv preprint arXiv:1802.05799\/} (2018).

\bibitem{silver2016mastering}
{\sc Silver, D., Huang, A., Maddison, C.~J., Guez, A., Sifre, L., Van
  Den~Driessche, G., Schrittwieser, J., Antonoglou, I., Panneershelvam, V.,
  Lanctot, M., et~al.}
\newblock Mastering the game of {Go} with deep neural networks and tree search.
\newblock {\em Nature 529}, 7587 (2016), 484--489.

\bibitem{silver2014deterministic}
{\sc Silver, D., Lever, G., Heess, N., Degris, T., Wierstra, D., and
  Riedmiller, M.}
\newblock Deterministic policy gradient algorithms.
\newblock In {\em ICML\/} (2014).

\bibitem{sutton1998reinforcement}
{\sc Sutton, R.~S., and Barto, A.~G.}
\newblock {\em Reinforcement Learning: An Introduction}.
\newblock MIT press Cambridge, 1998.

\bibitem{allreducealgs}
{\sc Thakur, R., Rabenseifner, R., and Gropp, W.}
\newblock Optimization of collective communication operations in {MPICH}.
\newblock {\em The International Journal of High Performance Computing
  Applications 19}, 1 (2005), 49--66.

\bibitem{tian2017elf}
{\sc Tian, Y., Gong, Q., Shang, W., Wu, Y., and Zitnick, C.~L.}
\newblock {ELF}: An extensive, lightweight and flexible research platform for
  real-time strategy games.
\newblock {\em Advances in Neural Information Processing Systems (NIPS)\/}
  (2017).

\bibitem{todorov2012mujoco}
{\sc Todorov, E., Erez, T., and Tassa, Y.}
\newblock Mujoco: A physics engine for model-based control.
\newblock In {\em Intelligent Robots and Systems (IROS), 2012 IEEE/RSJ
  International Conference on\/} (2012), IEEE, pp.~5026--5033.

\bibitem{van2010superhuman}
{\sc Van Den~Berg, J., Miller, S., Duckworth, D., Hu, H., Wan, A., Fu, X.-Y.,
  Goldberg, K., and Abbeel, P.}
\newblock Superhuman performance of surgical tasks by robots using iterative
  learning from human-guided demonstrations.
\newblock In {\em Robotics and Automation (ICRA), 2010 IEEE International
  Conference on\/} (2010), IEEE, pp.~2074--2081.

\bibitem{chain-replication}
{\sc van Renesse, R., and Schneider, F.~B.}
\newblock Chain replication for supporting high throughput and availability.
\newblock In {\em Proceedings of the 6th Conference on Symposium on Opearting
  Systems Design \& Implementation - Volume 6\/} (Berkeley, CA, USA, 2004),
  OSDI'04, USENIX Association.

\bibitem{drizzle}
{\sc Venkataraman, S., Panda, A., Ousterhout, K., Ghodsi, A., Armbrust, M.,
  Recht, B., Franklin, M., and Stoica, I.}
\newblock Drizzle: Fast and adaptable stream processing at scale.
\newblock In {\em Proceedings of the Twenty-Sixth ACM Symposium on Operating
  Systems Principles\/} (2017), SOSP '17, ACM.

\bibitem{hadoop}
{\sc White, T.}
\newblock {\em Hadoop: The Definitive Guide}.
\newblock O'Reilly Media, Inc., 2012.

\bibitem{spark-nsdi12}
{\sc Zaharia, M., Chowdhury, M., Das, T., Dave, A., Ma, J., McCauley, M.,
  Franklin, M.~J., Shenker, S., and Stoica, I.}
\newblock Resilient distributed datasets: A fault-tolerant abstraction for
  in-memory cluster computing.
\newblock In {\em Proceedings of the 9th USENIX conference on Networked Systems
  Design and Implementation\/} (2012), USENIX Association, pp.~2--2.

\bibitem{spark-cacm16}
{\sc Zaharia, M., Xin, R.~S., Wendell, P., Das, T., Armbrust, M., Dave, A.,
  Meng, X., Rosen, J., Venkataraman, S., Franklin, M.~J., Ghodsi, A., Gonzalez,
  J., Shenker, S., and Stoica, I.}
\newblock Apache {Spark}: A unified engine for big data processing.
\newblock {\em Commun. ACM 59}, 11 (Oct. 2016), 56--65.

\end{thebibliography}

\end{document}